\documentclass{article}
\usepackage{amssymb}
\usepackage{amsfonts}
\usepackage{amsmath}

\input{tcilatex}

\begin{document}

\begin{center}
\bigskip

{\large A MODIFIED Y-M ACTION WITH THREE\ FAMILIES\ OF FERMIONIC SOLITONS
AND PERTURBATIVE CONFINEMENT}

{\large \ }\bigskip

C. N. RAGIADAKOS

Pedagogical Institute

Mesogion 396, Agia Paraskevi, TK 15341, Greece

email: ragiadak@hol.gr, crag@pi-schools.gr

\bigskip

\textbf{ABSTRACT}
\end{center}

\begin{quote}
The dynamics of a four dimensional generally covariant modified SU(N)
Yang-Mills action, which depends on the complex structure of spacetime and
not its metric, is studied. A general solution of the complex structure
integrability conditions is found in the context of the $G_{2,2}$
Grassmannian manifold, which admits a global SL(4,C) symmetry group. A
convenient definition of the physical energy and momentum permits the study
of the vacuum and soliton sectors. The model has a set of conformally
SU(2,2) invariant vacua and a set of Poincar\'{e} invariant vacua. An
algebraic integrability condition of the complex structure classifies the
solitonic surfaces into three classes (families). The first class
(spacetimes with two principal null directions) contains the Kerr-Newman
complex structure, which has fermionic (electron-like) properties. That is
the correct fermionic gyromagnetic ratio (g=2) and it satisfies the correct
electron equations of motion. The conjugate complex structure determines the
antisoliton, which has the same mass and opposite charge. The fermionic
solitons are differentiated from the complex structure bosonic modes by the
periodicity condition on compactified spacetime. The non-periodicity of the
found solitonic\ complex structures is proved. The modification of the
Yang-Mills action has an essential consequence to the classical potential.
It generates a linear static\ potential instead of the Coulomb-like $\frac{1%
}{r}$ potential of the ordinary Yang-Mills action. This linear potential
implies that for every pure geometric soliton there are N solitonic gauge
field\ excitations, which are perturbatively confined. The present model
advocates a solitonic unification scheme without supersymmetry and/or
superstrings.\newpage
\end{quote}

{\LARGE Contents}

\textbf{1. INTRODUCTION}

\textbf{2. ACTION\ OF\ THE\ MODEL}

\qquad 2.1 Tetrad form of the action

\qquad 2.2 Examples of complex structures

\textbf{3. THE }$G_{2,2}$\textbf{\ GRASSMANNIAN\ MANIFOLDS}

\qquad 3.1 Bounded and unbounded realizations of $SU(2,2)$ classical domain

\qquad 3.2 Complex structures in $G_{2,2}$ context

\qquad 3.3 Induced metrics on spacetimes

\textbf{4. VACUA\ AND\ EXCITATION MODES}

\qquad 4.1 Physical energy-momentum

\qquad 4.2 Conformal and Poincar\'{e} vacua

\textbf{5. \textquotedblleft LEPTONIC\textquotedblright\ SOLITONS}

\qquad 5.1 Three families of solitons

\qquad 5.2 Massive complex structures of the 1$^{st}$ family

\qquad 5.3 Solitonic features of the massive structures

\qquad 5.4 Hopf invariants of complex structures

\qquad 5.5 Massless complex structures of the 1$^{st}$ family

\qquad 5.6 The 2$^{nd}$ and 3$^{rd}$\ family solitons may be unstable

\textbf{6. \textquotedblleft HADRONIC\textquotedblright\ SOLITONS\ AND\
CONFINEMENT}

\qquad \newpage

\renewcommand{\theequation}{\arabic{section}.\arabic{equation}}

\section{INTRODUCTION}

\setcounter{equation}{0}

Standard Model appears to be amazingly successful in all its experimental
testings. These successes can be found in many recent books in Quantum Field
Theory. But it is generally believed that it is not a complete theory
because it contains too many independent parameters. On the other hand many
apparent phenomena have not yet been proven or successfully described. Quark
confinement, the three generations of leptons, the corresponding three
generations of quarks and the apparent correspondence between leptons and
quarks are some characteristic physical phenomena, which have not yet been
understood in the context of Quantum Field Theory. These characteristic
features are proved to occur in the present slightly modified generally
covariant Yang-Mills model, which has fermionic solitons without fermionic
fields in its action.

General Relativity is actually a well established macroscopic theory. It is
based on the Einstein equation

\begin{equation}
E^{\mu \nu }\equiv R^{\mu \nu }-\frac{1}{2}\,R\,g^{\mu \nu }=8\pi k\,T^{\mu
\nu }  \label{1a1}
\end{equation}%
where $T^{\mu \nu }$ is the energy momentum tensor of the matter fields. $%
T^{\mu \nu }$ is an external non-geometric quantity, which is formally
imposed by hand. The classical mathematical problem of this theory is to
compute the metric tensor $g^{\mu \nu }$, which satisfies the Einstein
equation. The external character of $T^{\mu \nu }$ has always been
considered as a drawback of the theory and many efforts have been undertaken
to derive it from geometry.

The two successful mainstreams (Quantum Field Theory and General Relativity)
have been developed independently. Each branch has tried to incorporate the
other one without apparent success. The straightforward \textquotedblleft
covariantization\textquotedblright\ of the Standard Model action with the
Einstein gravitational term is not renormalizable. That is, it is not a self
consistent Quantum Field Theory. These failures led researchers to look for
non-conventional Lagrangian models. The mainstream of research turned first
into Supergravity without success and after into Superstrings, without
apparent experimental tests up to now. The first tests on supersymmetry are
expected to be provided by the Large Hadron Collider (LHC) experiments. If
supersymmetry is not found then research has to turn to more conventional
models like the present solitonic one.

General Relativity researchers tried to generate particles in the context of
Geometrodynamics, where matter is considered as a manifestation of geometry.
The fundamental idea and expectation is to derive all particles from pure
geometric quantities. The Einstein equation (\ref{1a1}) is seen as the
definition of the energy-momentum tensor of these particles. The
Einstein-Infeld-Hoffman theory\cite{E-I-H1938} of motion in General
Relativity may be considered the origin of the geometrodynamic ideas. In
this context the particle appears as a singularity of the Einstein tensor $%
E^{\mu \nu }$ and the equation of motion is derived from the
self-consistency identity $E^{\mu \nu };\nu \equiv 0$ and the definitions of
center of mass and the momenta. This result came up as a surprise for many
researchers, who were used to the linear character of the Maxwell equations,
and the problem has been extensively studied\cite{INFELD1957}. Apparently
the geometrodynamic point of view has also to postulate the (fundamental)
equations, which the solitonic manifold must satisfy, and how the well known
form of the Maxwell energy-momentum tensor is derived. In the Misner-Wheeler
model\cite{M-W1957} the fundamental equations are the Rainich conditions\cite%
{RAIN1925}. The metric configuration is the \textquotedblleft
particle\textquotedblright\ itself, and the electric charge is defined using
an Einstein-Rosen wormhole\cite{E-R1935} structure of space time. Despite
its great philosophical appeal this program fails to describe many
characteristic physical phenomena. It cannot explain the weak and the strong
interactions, its solitonic particles have continuous mass and charge
parameters, it is full of spinless solitons which do not appear in nature,
etc, etc. One may think that the Rainich conditions are responsible for
these unphysical consequences. But it has been proven that it is not the
case. The Finkelstein-Misner topological analysis\cite{F-M1959} showed that
any set of (fundamental) equations applied to the metric tensor cannot imply
soliton (particle, mass, etc.) discretization. It was this no-go result,
that blocked any investigation for a new realistic geometrodynamic model and
made this interesting idea to fade away.

Geometrodynamics has also failed to generate quantum phenomena but this
effort had two extraordinary results which are purely microscopic. The first
result is the derivation of the electron equations of motion\cite{PAP1951},%
\cite{CHASE1954},\cite{N-S2006} with the right terms without any effort of
model building. The second one is the observation\cite{CART1968},\cite%
{NEWM1973} that the Kerr-Newman spacetime has the electron gyromagnetic
ratio $g=2$. These two results strongly suggested the identification of the
electron with the charged Kerr manifold but the appropriate Quantum Field
Theoretic model was missing. The value of the present model is that it may
play this role or it may show the way how to find such a theory, which could
incorporate these extraordinary results into Quantum Field Theory. This
unification procedure does not need supersymmetry, because the fermionic
particles appear as solitons of the model. That is the proposed particle
unification scheme is of solitonic and not supersymmetric origin. In a
solitonic unification schemes Standard Model is simply an effective action
like the phonon actions in solids and fluids\cite{VOLO2001}. If the next few
years the LHC experiments do not find supersymmetric particles, we have to
turn to solitonic unification schemes as suggested by the present model.

The model started\cite{RAG1988} as a simple exercise to find a four
dimensional generally covariant action which would depend on the complex
structure of the spacetime and not on its metric. Recall that this property
characterizes the two dimensional string action. The purpose of this search
was to find a renormalizable generally covariant action without higher order
derivatives. Metric independence assures renormalizability\cite{RAG2008},
because the regularization procedure cannot generate non-renormalizable
geometric terms. Only topological anomalies may appear. Calculations\cite%
{RAG2008} of the first order one-loop diagrams in a convenient gauge
condition show that they are finite. The action\cite{RAG1990} of the model
is reviewed in section 2, where the properties of the Lorentzian complex
structure\cite{FLAHE1974} are reviewed\cite{FLAHE1976}. The new result of
this section is the general solution of the complex structure integrability
conditions using structure coordinates. A large part of the present paper is
devoted to review the mathematical background because it is no used in
current particle physics.

In section 3, the formalism of the Grassmannian manifolds and the classical
domains is applied to reveal the invariance of the complex structures under
the four dimensional global $SL(4,C)$ which is analogous to the $SL(2,C)$
symmetry of the string action. This mathematical background is necessary for
the reader to understand the vacua and soliton sectors of the model and how
global $SL(4,C)$ breaks down to the conformal $SU(2,2)$ and the physical
Poincar\'{e} symmetries, which are studied in section 4. The natural
emergence of the Poincar\'{e} group is the most interesting result of the
present model. It permits to find massive and massless stationary
axisymmetric solitons and to classify the complex structures using the Hopf
invariant. This rich physical content of the model is revealed in section 4.
The complex structures with solitonic properties are classified\cite{RAG1999}
into three classes relative to the number of sheets of the complex
structure. The first two-valued class is extensively studied. They are
solitons because their complex structures cannot be compactified. The
antisolitons are simply the complex conjugate complex structures of the
solitons. Solitons and antisolitons\ have the same mass but opposite
charges. The general forms of these massive and massless stationary
axisymmetric solitons are computed. An analogous calculation indicates that
the other two classes of solitons (with three and four sheets) do not
contain stable massive solitons.

The model contains only a Yang-Mills field and the ordinary (null) tetrad
which determines the Lorentzian complex structure. The symmetries of the
model do not permit the existence of fermionic fields. In section 5 we show
that the modification of the Yang-Mills action, which makes it independent
of the metric tensor, has a characteristic physical consequence. The static
potential of a source is no longer $\frac{1}{r}$ but it is linear, which
could confine the \textquotedblleft colored\textquotedblright\ sources\cite%
{RAG1999}. From the two dimensional solitonic models we know that the
solitons may be excited by the field modes. Analogous excitations are
expected in the present case too. That is, the solitonic complex structures
of the model may be excited by the gauge field modes. Then these excited
solitons are perturbatively confined because of the linear gauge field\
potential. Only \textquotedblleft colorless\textquotedblright\ states may
exist free, which is a characteristic property of the hadrons. Notice that
this confinement mechanism implies a strict correspondence between
\textquotedblleft leptonic\textquotedblright\ pure geometric\ solitons with
vanishing gauge field \ and the \textquotedblleft
hadronic\textquotedblright\ ones with non-vanishing gauge field.

\section{ACTION\ OF\ THE\ MODEL}

\setcounter{equation}{0}

The ordinary (Euclidean) almost complex structure is a \underline{real}
tensor $J_{\mu }^{\;\nu }$, normalized by the condition 
\begin{equation}
J_{\mu }^{\;\rho }J_{\rho }^{\;\nu }=-\delta _{\mu }^{\nu }  \label{2a1}
\end{equation}%
It defines an (integrable) complex structure, if it satisfies the Nijenhuis
integrability condition

\begin{equation}
J_{\mu }^{\;\sigma }\left( \partial _{\sigma }J_{\rho }^{\;\nu }-\partial
_{\rho }J_{\sigma }^{\;\nu }\right) -J_{\rho }^{\;\sigma }\left( \partial
_{\sigma }J_{\mu }^{\;\nu }-\partial _{\mu }J_{\sigma }^{\;\nu }\right) =0
\label{2a2}
\end{equation}%
Then the manifold over which $J_{\mu }^{\;\nu }$ exists, becomes a complex
manifold.

A complex structure is compatible with the metric tensor $g_{\mu \nu }$ of
the manifold, if the two tensors satisfy the relation 
\begin{equation}
J_{\rho }^{\;\mu }J_{\sigma }^{\;\nu }g_{\mu \nu }=g_{\rho \sigma }
\label{2a3}
\end{equation}%
at any point of the manifold. If the signature of spacetime is Lorentzian,
there is always a coordinate transformation such that the metric tensor
takes the form of the Minkowski metric $\eta _{\mu \nu }$ at a given point.
Then we see that the real tensor $J_{\mu }^{\;\nu }$ defines a Lorentz
transformation at the given point. However there is no real Lorentz
transformation, which satisfies the normalization condition (\ref{2a1}) of
the complex structure. Notice that this incompatibility is a pure local
property and it is not related to the global structure of spacetime.

Hence the Lorentzian signature of spacetime is not compatible with a 
\underline{real} tensor (complex structure) $J_{\mu }^{\;\nu }$. The notion
of the Lorentzian complex structure has been generalized\cite{FLAHE1974} to
include complex tensors $J_{\mu }^{\;\nu }$. I anticipate that the existence
of antisolitons in the present model is based on this particular property of
the Lorentzian complex structure. This (modified) complex structure has been
extensively studied by Flaherty\cite{FLAHE1976}. It can be shown that there
is always a null tetrad $(\ell _{\mu },\,n_{\mu },\,m_{\mu },\,\overline{m}%
_{\mu })$ such that the metric tensor and the complex structure tensor take
the form

\begin{equation}
\begin{array}{l}
g_{\mu \nu }=\ell _{\mu }n_{\nu }+n_{\mu }\ell _{\nu }-m_{{}\mu }\overline{m}%
_{\nu }-\overline{m}_{\mu }m_{\nu } \\ 
\\ 
J_{\mu }^{\;\nu }=i(\ell _{\mu }n^{\nu }-n_{\mu }\ell ^{\nu }-m_{\mu }%
\overline{m}^{\nu }+\overline{m}_{\mu }m^{\nu })%
\end{array}
\label{2a4}
\end{equation}%
The integrability condition of this complex structure implies the Frobenius
integrability conditions of the pairs $(\ell _{\mu },\,\,m_{\mu })$ and $%
(n_{\mu },\,\overline{m}_{\mu })$. That is

\begin{equation}
\begin{array}{l}
(\ell ^{\mu }m^{\nu }-\ell ^{\nu }m^{\mu })(\partial _{\mu }\ell _{\nu
})=0\;\;\;\;,\;\;\;\;(\ell ^{\mu }m^{\nu }-\ell ^{\nu }m^{\mu })(\partial
_{\mu }m_{\nu })=0 \\ 
\\ 
(n^{\mu }m^{\nu }-n^{\nu }m^{\mu })(\partial _{\mu }n_{\nu
})=0\;\;\;\;,\;\;\;\;(n^{\mu }m^{\nu }-n^{\nu }m^{\mu })(\partial _{\mu
}m_{\nu })=0%
\end{array}
\label{2a5}
\end{equation}

Frobenius theorem states that there are four complex functions $(z^{\alpha
},\;z^{\widetilde{\alpha }})$,\ $\alpha =0,\ 1$ , such that

\begin{equation}
dz^{\alpha }=f_{\alpha }\ \ell _{\mu }dx^{\mu }+h_{\alpha }\ m_{\mu }dx^{\mu
}\;\;\;\;,\;\;\;dz^{\widetilde{\alpha }}=f_{\widetilde{\alpha }}\ n_{\mu
}dx^{\mu }+h_{\widetilde{\alpha }}\ \overline{m}_{\mu }dx^{\mu }\;
\label{2a6}
\end{equation}%
These four functions are the structure coordinates of the (integrable)
complex structure. Notice that in the present case of Lorentzian spacetimes
the coordinates $z^{\widetilde{\alpha }}$ are not complex conjugate of $%
z^{\alpha }$, because $J_{\mu }^{\;\nu }$ is no longer a real tensor.

The reality conditions of the Newman-Penrose null tetrad $(\ell _{\mu
},\,n_{\mu },\,m_{\mu },\,\overline{m}_{\mu })$ imply

\begin{equation}
\begin{array}{l}
dz^{0}\wedge dz^{1}\wedge d\overline{z^{0}}\wedge d\overline{z^{1}}=0 \\ 
\\ 
dz^{\widetilde{0}}\wedge dz^{\widetilde{0}}\wedge d\overline{z^{0}}\wedge d%
\overline{z^{1}}=0 \\ 
\\ 
dz^{\widetilde{0}}\wedge dz^{\widetilde{0}}\wedge d\overline{z^{\widetilde{0}%
}}\wedge d\overline{z^{\widetilde{0}}}=0%
\end{array}
\label{2a8}
\end{equation}%
These relations are directly proven after a substitution of (\ref{2a6}).
They are equivalent to the existence of two real functions $\Omega _{0}$ , $%
\Omega _{\widetilde{0}}$ and a complex one $\Omega $, such that

\begin{equation}
\Omega _{0}\left( z^{\alpha },\overline{z^{\alpha }}\right) =0\quad ,\quad
\Omega \left( z^{\widetilde{\alpha }},\overline{z^{\alpha }}\right) =0\quad
,\quad \Omega _{\widetilde{0}}\left( z^{\widetilde{\alpha }},\overline{z^{%
\widetilde{\alpha }}}\right) =0  \label{2a9}
\end{equation}%
Notice that these relations provide an algebraic solution to the problem of
complex structures on a spacetime. They are much easier handled than the
PDEs (\ref{2a5}). In the next section they will be transcribed in the $%
G_{2,2}$ Grassmannian manifold context providing a powerful mathematical
machinery for the computation of complex structures.

The integrability conditions of the complex structure can be formulated in
the spinor formalism. They imply that both spinors $o^{A}$ and $\iota ^{A}$
of the dyad satisfy the same PDE

\begin{equation}
\xi ^{A}\xi ^{B}\nabla _{AA^{\prime }}\ \xi _{B}=0  \label{2a10}
\end{equation}%
where $\nabla _{AA^{\prime }}$ is the covariant derivative connected to the
vierbein $e_{a}^{\ \mu }$. This relation is equivalent to the existence of a
complex vector field $\tau ^{A^{\prime }B}$ such that

\begin{equation}
\nabla _{(A^{\prime }}^{A^{\prime }}\ \xi _{B)}=\tau _{(A}^{A^{\prime }}\xi
_{B)}  \label{2a11}
\end{equation}%
Using the relation\cite{P-R1984}

\begin{equation}
\nabla _{A^{\prime }(A}\nabla _{B}^{A^{\prime }}\ \xi _{C)}=\Psi _{ABCD}\xi
^{D}  \label{2a12}
\end{equation}%
one can show that both $o^{A}$ and $\iota ^{A}$ satisfy the algebraic
integrability condition

\begin{equation}
\Psi _{ABCD}\xi ^{A}\xi ^{B}\xi ^{C}\xi ^{D}=0  \label{2a13}
\end{equation}%
Namely, they are principal directions of the Weyl spinor $\Psi _{ABCD}$.
Therefore a curved spacetime may admit a limited number of complex
structures, which are directly related to its principal null directions. If
the Weyl curvature vanishes, there is no restriction on the proper spinor
basis. In this case the manifold is conformally flat and the integrability
conditions are completely solved via Kerr's theorem\cite{FLAHE1976}.

\subsection{Tetrad and structure coordinate forms of the action}

The string action describes the dynamics of 2-dimensional surfaces in a
multidimensional space. Its form%
\begin{equation}
I_{S}=\frac{1}{2}\int d^{2}\!\xi \ \sqrt{-\gamma }\ \gamma ^{\alpha \beta }\
\partial _{\alpha }X^{\mu }\partial _{\beta }X^{\nu }\eta _{\mu \nu }
\label{2b1}
\end{equation}%
does not essentially depend on the metric $\gamma ^{\alpha \beta }$ of the
2-dimensional surface. It depends on its structure coordinates $(z^{0},\ z^{%
\widetilde{0}})$, because in these coordinates it takes the metric
independent form%
\begin{equation}
I_{S}=\int d^{2}\!z\ \partial _{0}X^{\mu }\partial _{\widetilde{0}}X^{\nu
}\eta _{\mu \nu }  \label{2b2}
\end{equation}%
All the wonderful properties of the string model are essentially based on
this characteristic feature of the string action.

The plausible question\cite{RAG1988} and exercise is \textquotedblleft what
4-dimensional action with first order derivatives depends on the complex
structure but it does not depend on the metric of the
spacetime?\textquotedblright . The additional expectation is that such an
action may be formally renormalizable because the regularization procedure
will not generate geometric counterterms. The term \textquotedblleft
formally\textquotedblright\ is used because the 4-dimensional action may
have anomalies which could destroy renormalizability, as it happens in the
string action. Recall that the string and superstring actions are
self-consistent only in precise dimensions, where the cancellation of the
anomaly occurs.

A four dimensional action which satisfies the above criterion was found. The
null tetrad form of this action\cite{RAG1990} of the present model is 
\begin{equation}
\begin{array}{l}
I_{G}=\int d^{4}\!x\ \sqrt{-g}\ \left\{ \left( \ell ^{\mu }m^{\rho
}F_{\!j\mu \rho }\right) \left( n^{\nu }\overline{m}^{\sigma }F_{\!j\nu
\sigma }\right) +\left( \ell ^{\mu }\overline{m}^{\rho }F_{\!j\mu \rho
}\right) \left( n^{\nu }m^{\sigma }F_{\!j\nu \sigma }\right) \right\} \\ 
\\ 
F_{j\mu \nu }=\partial _{\mu }A_{j\nu }-\partial _{\nu }A_{j\mu }-\gamma
\,f_{jik}A_{i\mu }A_{k\nu }%
\end{array}
\label{2b3}
\end{equation}%
where $A_{j\mu }$ is a gauge field and $(\ell _{\mu },\,n_{\mu },\,m_{\mu
},\,\overline{m}_{\mu })$ is an integrable null tetrad. The difference
between the present action and the ordinary Yang-Mills action becomes more
clear in the following form of the action. 
\begin{equation}
I_{G}=-\frac{1}{8}\int d^{4}\!x\ \sqrt{-g}\ \left( 2g^{\mu \nu }\ g^{\rho
\sigma }-J^{\mu \nu }\ J^{\rho \sigma }-\overline{J^{\mu \nu }}\ \overline{%
J^{\rho \sigma }}\right) F_{\!j\mu \rho }F_{\!j\nu \sigma }  \label{2b4}
\end{equation}%
where $g_{\mu \nu }$ is a metric derived from the null tetrad and $J_{\mu
}^{\;\nu }$ is the tensor of the integrable complex structure.

Like the 2-dimensional string action, the metric independence of the present
action appears when we transcribe it in its structure coordinates form 
\begin{equation}
\begin{array}{l}
I_{G}=\int d^{4}\!z\ F_{\!j01}F_{\!j\widetilde{0}\widetilde{1}}+comp.\ conj.
\\ 
\\ 
F_{j_{ab}}=\partial _{a}A_{jb}-\partial _{a}A_{jb}-\gamma
\,f_{jik}A_{ia}A_{kb}%
\end{array}
\label{2b5}
\end{equation}%
This transcription is possible because the metric and the integrable null
tetrad take simple forms in the structure coordinates system.

In the case of the string action we do not need additional conditions
because any orientable 2-dimensional surface admits a complex structure. But
in the case of 4-dimensional surfaces, the integrability of the complex
structure has to be imposed through precise conditions. These integrability
conditions may be imposed either on the tetrad (\ref{2a5}) or on the
structure coordinates (\ref{2a8}), using the ordinary procedure of Lagrange
multipliers. These different possibilities will provide the various forms of
the action which are equivalent, at least in the classical level. Its
different variations should be seen as different ways to write down the
integration measure over the complex structures of the 4-dimensional
Lorentzian manifolds. The additional action term with the integrability
conditions on the null tetrad is 
\begin{equation}
\begin{array}{l}
I_{C}=-\int d^{4}\!x\ \{\phi _{0}(\ell ^{\mu }m^{\nu }-\ell ^{\nu }m^{\mu
})(\partial _{\mu }\ell _{\nu })+ \\ 
\\ 
\qquad +\phi _{1}(\ell ^{\mu }m^{\nu }-\ell ^{\nu }m^{\mu })(\partial _{\mu
}m_{\nu })+\phi _{\widetilde{0}}(n^{\mu }\overline{m}^{\nu }-n^{\nu }%
\overline{m}^{\mu })(\partial _{\mu }n_{\nu })+ \\ 
\\ 
\qquad +\phi _{\widetilde{1}}(n^{\mu }\overline{m}^{\nu }-n^{\nu }\overline{m%
}^{\mu })(\partial _{\mu }\overline{m}_{\nu })+c.conj.\}%
\end{array}
\label{2b6}
\end{equation}%
This Lagrange multiplier makes the complete action $I=I_{G}+I_{C}$
self-consistent and the usual quantization techniques may be used\cite%
{RAG1992}.

The local symmetries of the action are a) the well known local gauge
transformations, b) the reparametrization symmetry as it is the case in any
generally covariant action and c) the following extended Weyl transformation
of the tetrad 
\begin{equation}
\begin{array}{l}
\ell _{\mu }^{\prime }=\chi _{1}\ell _{\mu }\quad ,\quad n_{\mu }^{\prime
}=\chi _{2}n_{\mu }\quad ,\quad m_{\mu }^{\prime }=\chi m_{\mu } \\ 
\\ 
\phi _{0}^{\prime }=\phi _{0}\frac{\chi _{2}\overline{\chi }}{\chi _{1}}%
\quad ,\quad \phi _{1}^{\prime }=\phi _{1}\frac{\chi _{2}\overline{\chi }}{%
\chi } \\ 
\\ 
\phi _{\widetilde{0}}^{\prime }=\phi _{\widetilde{0}}\frac{\chi _{1}\chi }{%
\chi _{2}}\quad ,\quad \phi _{\widetilde{1}}^{\prime }=\phi _{\widetilde{1}}%
\frac{\chi _{1}\chi }{\overline{\chi }} \\ 
\\ 
g^{\prime }=g(\chi _{1}\chi _{2}\chi \overline{\chi })^{2}%
\end{array}
\label{2b7}
\end{equation}%
where $\chi _{1},\chi _{2}$ are real functions and $\chi $ is a complex one.

\subsection{Examples of complex structures}

In order to make a selfconsistent paper we will present here some examples
of complex structures, which can also be found in the works of Flaherty. The
configurations of these complex structures will be used in the next sections.

The spinorial form of the integrability condition of the complex structure
is conformally invariant. It is invariant under a spinor $\xi ^{A}$
multiplication with an arbitrary function, therefore we do not loose
generality assuming the form $\xi ^{A}=[1,\ \lambda ]$. Then, in the
Cartesian coordinates of a conformally flat spacetime the spinorial
integrability conditions become the Kerr differential equations 
\begin{equation}
\begin{array}{l}
\lambda ^{A}\lambda ^{B}\nabla _{A^{\prime }A}\lambda _{B}=0\quad
\Longleftrightarrow \\ 
\\ 
(\partial _{0^{\prime }0}\lambda )+\lambda (\partial _{0^{\prime }1}\lambda
)=0\ \ and\ \ (\partial _{1^{\prime }0}\lambda )+\lambda (\partial
_{1^{\prime }1}\lambda )=0%
\end{array}
\label{2d1}
\end{equation}%
where the Penrose spinorial notation is used with%
\begin{equation}
\begin{array}{l}
x^{A^{\prime }A}=x^{\mu }\sigma _{\mu }^{A^{\prime }A}=%
\begin{pmatrix}
x^{0}+x^{3} & (x^{1}+ix^{2}) \\ 
(x^{1}-ix^{2}) & x^{0}-x^{3}%
\end{pmatrix}
\\ 
\\ 
x_{A^{\prime }A}=%
\begin{pmatrix}
x^{0}-x^{3} & -(x^{1}-ix^{2}) \\ 
-(x^{1}+ix^{2}) & x^{0}+x^{3}%
\end{pmatrix}
\\ 
\\ 
\partial _{A^{\prime }A}=\frac{\partial }{\partial x^{A^{\prime }A}}=\sigma
_{A^{\prime }A}^{\mu }\partial _{\mu }=%
\begin{pmatrix}
\partial _{0}+\partial _{3} & \partial _{1}-i\partial _{2} \\ 
\partial _{1}+i\partial _{2} & \partial _{0}-\partial _{3}%
\end{pmatrix}
\\ 
\end{array}
\label{2d01}
\end{equation}

Kerr's theorem states that a general solution of these equations is any
function $\lambda (x^{A^{\prime }B})$, which satisfies a relation of the form%
\begin{equation}
K(\lambda ,\ x_{0^{\prime }0}+x_{0^{\prime }1}\lambda ,\ x_{1^{\prime
}0}+x_{1^{\prime }1}\lambda )=0  \label{2d2}
\end{equation}%
where $K(\cdot ,\ \cdot ,\ \cdot )$\ is an arbitrary function.

Notice that in a conformally flat spacetime, the two solutions $\lambda _{1}$
and $\lambda _{2}$, which determine the spinor dyad $o^{A}\propto (1,\
\lambda _{1})$\ and $\iota ^{A}\propto (1,\ \lambda _{2})$,\ completely
decouple. A characteristic example of a Minkowski spacetime complex
structure is given by the two solutions of the quadratic (Kerr) function%
\begin{equation}
(x-iy)\lambda ^{2}+2(z-ia)\lambda -(x+iy)=0  \label{2d3}
\end{equation}%
where the ordinary Cartesian coordinates $x^{1}=x,\ x^{2}=y,\ x^{3}=z$ are
used. This Kerr function is time independent and determines a static complex
structure. The two solutions are

\begin{equation}
\lambda _{1,2}=\frac{-(z-ia)\pm \sqrt{\Delta }}{x-iy}\quad ,\quad \Delta
=x^{2}+y^{2}+z^{2}-a^{2}-2iaz  \label{2d4}
\end{equation}

The corresponding spinor basis (dyad) is%
\begin{equation}
o^{A}=[1\ ,\ \frac{-(z-ia)+\sqrt{\Delta }}{x-iy}]\quad ,\quad \iota ^{A}=-%
\frac{x-iy}{2\sqrt{\Delta }}[1,\ \frac{-(z-ia)-\sqrt{\Delta }}{x-iy}]
\label{2d5}
\end{equation}%
The corresponding null tetrad is%
\begin{equation}
\begin{array}{l}
L\propto \left[ (1+\lambda _{1}\overline{\lambda _{1}})dt-(\lambda _{1}+%
\overline{\lambda _{1}})dx-i(\overline{\lambda _{1}}-\lambda
_{1})dy-(1-\lambda _{1}\overline{\lambda _{1}})dz\right] \\ 
\\ 
M\propto \left[ (1+\lambda _{1}\overline{\lambda _{2}})dt-(\lambda _{1}+%
\overline{\lambda _{2}})dx-i(\overline{\lambda _{2}}-\lambda
_{1})dy-(1-\lambda _{1}\overline{\lambda _{2}})dz\right] \\ 
\\ 
N\propto \left[ (1+\lambda _{2}\overline{\lambda _{2}})dt-(\lambda _{2}+%
\overline{\lambda _{2}})dx-i(\overline{\lambda _{2}}-\lambda
_{2})dy-(1-\lambda _{2}\overline{\lambda _{2}})dz\right]%
\end{array}
\label{2d6}
\end{equation}%
which is the \textquotedblleft flatprint\textquotedblright\ null tetrad of
the Kerr-Newman manifold. In the case of $a=0$ it becomes the trivial
\textquotedblleft spherical\textquotedblright\ complex structure.

In the case of conformally flat spacetimes the structure coordinates $%
z^{\alpha }$ are two independent functions of $(\lambda _{1},\ x_{0^{\prime
}0}+x_{0^{\prime }1}\lambda _{1},\ x_{1^{\prime }0}+x_{1^{\prime }1}\lambda
_{1})$ and the structure coordinates $z^{\widetilde{\alpha }}$ are
respectively two independent functions of $(\lambda _{2},\ x_{0^{\prime
}0}+x_{0^{\prime }1}\lambda _{2},\ x_{1^{\prime }0}+x_{1^{\prime }1}\lambda
_{2})$. It is convenient to use the following structure coordinates%
\begin{equation}
\begin{array}{l}
z^{0}=t-\sqrt{\Delta }-ia\quad ,\quad z^{1}=\frac{-(z-ia)+\sqrt{\Delta }}{%
x-iy} \\ 
\\ 
z^{\widetilde{0}}=t+z+\frac{x^{2}+y^{2}}{z+\sqrt{\Delta }-ia}\quad ,\quad z^{%
\widetilde{1}}=-\frac{x-iy}{z+\sqrt{\Delta }-ia}%
\end{array}
\label{2d7}
\end{equation}

Notice that this complex structure cannot be defined over the whole
Minkowski spacetime, because it is singular when $o^{A}\propto \iota ^{A}$,
which occurs at

\begin{equation}
\begin{array}{l}
z=0\quad ,\quad x^{2}+y^{2}=a^{2} \\ 
\end{array}
\label{2d8}
\end{equation}%
We will see below that these points do not belong to the Grassmannian
manifold. Using the Lindquist coordinates $(t,\ r,\ \theta ,\ \varphi )$

\begin{equation}
\begin{array}{l}
x=(r\cos \varphi +a\sin \varphi )\sin \theta \\ 
\\ 
y=(r\sin \varphi -a\cos \varphi )\sin \theta \\ 
\\ 
z=r\cos \theta%
\end{array}
\label{2d9}
\end{equation}%
the structure coordinates take the form%
\begin{equation}
\begin{array}{l}
z^{0}=t-r+ia\cos \theta -ia\quad ,\quad z^{1}=e^{i\varphi }\tan \frac{\theta 
}{2} \\ 
\\ 
z^{\widetilde{0}}=t+r-ia\cos \theta +ia\quad ,\quad z^{\widetilde{1}}=-\frac{%
r+ia}{r-ia}e^{-i\varphi }\tan \frac{\theta }{2}%
\end{array}
\label{2d10}
\end{equation}%
The Minkowski spacetime null tetrad takes the form%
\begin{equation}
\begin{array}{l}
L_{\mu }dx^{\mu }=dt-dr-a\sin ^{2}\theta \ d\varphi \\ 
\\ 
N_{\mu }dx^{\mu }=\frac{r^{2}+a^{2}}{2(r^{2}+a^{2}\cos ^{2}\theta )}[dt+%
\frac{r^{2}+2a^{2}\cos ^{2}\theta -a^{2}}{r^{2}+a^{2}}dr-a\sin ^{2}\theta \
d\varphi ] \\ 
\\ 
M_{\mu }dx^{\mu }=\frac{-1}{\sqrt{2}(r+ia\cos \theta )}[-ia\sin \theta \
(dt-dr)+(r^{2}+a^{2}\cos ^{2}\theta )d\theta + \\ 
\qquad \qquad +i\sin \theta (r^{2}+a^{2})d\varphi ]%
\end{array}
\label{2d13}
\end{equation}

A simple way\cite{RAG1991}$^{,}$\cite{RAG1999} to find a curved space
complex structure is the Kerr-Schild ansatz

\begin{equation}
\ell _{\mu }=L_{\mu }\quad ,\quad m_{\mu }=M_{\mu }\quad ,\quad n_{\mu
}=N_{\mu }+f(x)\ L_{\mu }  \label{2e1}
\end{equation}%
where the null tetrad $(L_{\mu },\ N_{\mu },\ M_{\mu },\ \overline{M}_{\mu
}) $ determines an integrable flat complex structure. In the case of the
static (\ref{2d13}) complex structure, $(\ell _{\mu },\ n_{\mu },\ m_{\mu
},\ \overline{m}_{\mu })$ is integrable for

\begin{equation}
f=\frac{h(r)}{2(r^{2}+a^{2}\cos ^{2}\theta )}  \label{2e2}
\end{equation}%
where $h(r)$ is an arbitrary function. Notice that for $h(r)=-2mr+e^{2}$ the
Kerr-Newman space-time is found. A set of structure coordinates of the
curved complex structure, which are smooth deformations of the Minkowski
complex structure, are

\begin{equation}
\begin{array}{l}
z^{0}=t-r+ia\cos \theta -ia\quad ,\quad z^{1}=e^{i\varphi }\tan \frac{\theta 
}{2} \\ 
\\ 
z^{\widetilde{0}}=t+r-ia\cos \theta +ia-2f_{1}\quad ,\quad z^{\widetilde{1}%
}=-\frac{r+ia}{r-ia}\ e^{2iaf_{2}}\ e^{-i\varphi }\tan \frac{\theta }{2}%
\end{array}
\label{2e3}
\end{equation}%
where the two new functions are

\begin{equation}
f_{1}(r)=\int \frac{h}{r^{2}+a^{2}+h}\ dr\quad ,\quad f_{2}(r)=\int \frac{h}{%
(r^{2}+a^{2}+h)(r^{2}+a^{2})}\ dr  \label{2e4}
\end{equation}

In the present model these configurations are seen as solitons. The complex
structure $J_{\mu }^{\;\rho }$ describes a fermionic soliton with charge $e$
and its complex conjugate $\overline{J_{\mu }^{\;\rho }}$ describes an
antisoliton with charge $-e$.\pagebreak

\section{THE G$_{2,2}$\ GRASSMANNIAN\ MANIFOLD}

\setcounter{equation}{0}

The present work is heavily based on projective spaces and the classical
domains, therefore a short review of the projective Grassmannian manifolds
and the $SU(2,2)$ classical domain is needed. The projective space $CP^{3}$
is the set of non vanishing 4-d complex vector $Z^{m}\ ,\ m=0,1,2,3$ with
the equivalence relation $X^{m}\sim Y^{m}$\ if there exists a non vanishing
complex number $c$ such that $X^{m}=cY^{m}$. Then the natural topology of $%
C^{4}$ induces a well defined topology in $CP^{3}$. The coordinates $Z^{m}$\
are called homogeneous coordinates and the three coordinates $y^{I}=[\frac{%
Z^{1}}{Z^{0}}\ ,\ $ $\frac{Z^{2}}{Z^{0}}\ ,\ \frac{Z^{3}}{Z^{0}}]$ are
called projective coordinates in the $Z^{0}\neq 0$ coordinate neighborhood.
Every two elements $X^{m1}$ and $X^{m2}$\ of $CP^{3}$ determine a $2\times 2$
matrix $r_{A^{\prime }B}$ such that%
\begin{equation}
X^{mi}=%
\begin{pmatrix}
\lambda ^{Ai} \\ 
-ir_{A^{\prime }B}\lambda ^{Bi}%
\end{pmatrix}
\label{4a1}
\end{equation}%
where they are written in the chiral representation. Penrose\cite{PEN1967}
has observed that a general solution of the Kerr theorem, which determines
the geodetic and shear free congruences, take the simple form $K(Z^{m})=0$,
where \ $K(Z^{m})$ is a homogeneous function.\ In the case of a first degree
polynomial $K(Z^{m})=S_{m}Z^{m}$ with $S_{m}=[S_{A},S^{B^{\prime }}]$ we may
define $\omega _{A}$%
\begin{equation}
\omega _{A}\lambda ^{A}=S_{m}Z^{m}=(S_{A}-iS^{B^{\prime }}r_{B^{\prime
}A})\lambda ^{A}  \label{4a2}
\end{equation}%
After a straightforward calculation we find that $\omega _{A}\equiv \rho
_{A}-i\tau ^{B^{\prime }}r_{B^{\prime }A}$ satisfy the differential equation%
\begin{equation}
\partial _{A^{\prime }(B}\omega _{C)}=\frac{\partial \ \omega _{C}}{\partial
r^{A^{\prime }B}}+\frac{\partial \ \omega _{B}}{\partial r^{A^{\prime }C}}=0
\label{4a3}
\end{equation}%
Penrose points out that the inverse is also true. The space of the solutions
of this differential equation is $CP^{3}$. He called this differential
equation \textquotedblleft twistor equation\textquotedblright\ and the
projective space $CP^{3}$ twistor space. One can easily show that if $X^{mi}$%
\ of $CP^{3}$ satisfy the relations%
\begin{equation}
X^{i\dag }EX^{j}=0\quad ,\quad \forall \ i,\ j  \label{4a4}
\end{equation}%
with 
\begin{equation}
E=%
\begin{pmatrix}
0 & I \\ 
I & 0%
\end{pmatrix}
\label{4a5}
\end{equation}%
the generally complex $2\times 2$ matrix $r_{A^{\prime }B}$ becomes
Hermitian and it transforms as the Cartesian coordinates of the Minkowski
spacetime under the Poincar\'{e} subgroup of the projective transformations $%
SL(4,C)$ of $CP^{3}$. Penrose tried to extract physical meaning from these
relations. I will not continue on twistor mode of thinking, because I want
to avoid confusion of the present conventional quantum field theoretic model
with the Penrose twistor program. But many times in the present work we will
use the twistor formalism and its spinor notation, because it is
computationally very effective.

In the case of a second degree polynomial $K(Z^{m})=S_{mn}Z^{m}Z^{n}$ with%
\begin{equation}
S_{mn}=%
\begin{pmatrix}
S_{AB} & S_{A}^{\quad B^{\prime }} \\ 
S_{\quad B}^{A^{\prime }} & S^{A^{\prime }B^{\prime }}%
\end{pmatrix}%
\quad where\quad S_{AB}=S_{BA}\ ,\ S_{\quad B}^{A^{\prime }}=S_{B}^{\quad
A^{\prime }}  \label{4a6}
\end{equation}%
we define the spinor $\omega _{AB}$ 
\begin{equation}
\omega _{AB}\lambda ^{A}\lambda
^{B}=S_{mn}Z^{m}Z^{n}=(S_{AB}-iS_{A}^{A^{\prime }}r_{A^{\prime
}B}-iS_{B}^{B^{\prime }}r_{B^{\prime }A}-S^{A^{\prime }B^{\prime
}}r_{A^{\prime }A}r_{B^{\prime }B})\lambda ^{A}\lambda ^{B}  \label{4a7}
\end{equation}%
which satisfies the relation%
\begin{equation}
\partial _{A^{\prime }(B}\omega _{CD)}=0  \label{4a8}
\end{equation}

In the case of a fourth degree polynomial $%
K(Z^{m})=S_{mnpq}Z^{m}Z^{n}Z^{p}Z^{q}$ we find%
\begin{equation}
\begin{array}{l}
\omega _{ABCD}=S_{ABCD}-iS_{(ACD}^{A^{\prime }}r_{A^{\prime
}B)}-S_{(CD}^{A^{\prime }B^{\prime }}r_{A^{\prime }A}r_{B^{\prime }B)}+ \\ 
\qquad +iS_{(D}^{A^{\prime }B^{\prime }C^{\prime }}r_{A^{\prime
}A}r_{B^{\prime }B}r_{C^{\prime }C)}+iS^{A^{\prime }B^{\prime }C^{\prime
}D^{\prime }}r_{A^{\prime }A}r_{B^{\prime }B}r_{C^{\prime }C}r_{C^{\prime }C}%
\end{array}
\label{4a9}
\end{equation}%
which also satisfies the twistor equation%
\begin{equation}
\partial _{A^{\prime }(B}\omega _{CDEF)}=0  \label{4a10}
\end{equation}

It is proved that a spinor $\lambda ^{A}$, which satisfies the fourth degree
homogeneous polynomial%
\begin{equation}
\omega _{ABCD}\lambda ^{A}\lambda ^{B}\lambda ^{C}\lambda ^{D}=0
\label{4a11}
\end{equation}%
determines a geodetic and shear free congruence in Minkowski spacetime. Two
roots of this polynomial define a complex structure. This relation is
useful, because it will coincide with the algebraic integrability condition
on the curved spacetime in the weak gravity approximation. That is, we
expect the Weyl spinorial tensor $\Psi _{ABCD}$ to become proportional with $%
\omega _{ABCD}$ in the weak gravity limit and in an appropriate (Cartesian)
coordinate system.

Let us now turn to the definition of the Grassmannian projective manifold $%
G_{2,2}$. Consider the set of the $4\times 2$ complex matrices of rank 2%
\begin{equation}
T=\left( 
\begin{array}{c}
T_{1} \\ 
T_{2}%
\end{array}%
\right)  \label{4b1}
\end{equation}%
with the equivalence relation $T\sim T^{\prime }$ if there exists a $2\times
2$ regular matrix $S$ such that%
\begin{equation}
T^{\prime }=TS  \label{4b2}
\end{equation}%
The coordinates%
\begin{equation}
z=T_{2}T_{1}^{-1}  \label{4b3}
\end{equation}%
completely determine the points of the set. The topology of the $4\times 2$
matrices implies a well defined topology in this projective manifold $%
G_{2,2} $. The coordinates $T$ are called homogeneous coordinates and the
coordinates $z$ are called projective coordinates. Under a general linear $%
4\times 4$ transformation%
\begin{equation}
\begin{pmatrix}
T_{1}^{\prime } \\ 
T_{2}^{\prime }%
\end{pmatrix}%
=%
\begin{pmatrix}
A_{11} & A_{12} \\ 
A_{21} & A_{22}%
\end{pmatrix}%
\left( 
\begin{array}{c}
T_{1} \\ 
T_{2}%
\end{array}%
\right)  \label{4b4}
\end{equation}%
the inhomogeneous coordinates transform as%
\begin{equation}
z^{\prime }=\left( A_{21}+A_{22}\ z\right) \left( A_{11}+A_{12}\ z\right)
^{-1}  \label{4b5}
\end{equation}%
which is called fractional transformation and it preserves the compact
manifold $G_{2,2}$, which is called Grassmannian manifold.

\subsection{B\textbf{ounded and unbounded realizations of the\ SU(2,2)
classical domain}}

The points of $G_{2,2}$ with positive definite $2\times 2$ matrix%
\begin{equation}
\begin{pmatrix}
T_{1}^{\dagger } & T_{2}^{\dagger }%
\end{pmatrix}%
\left( 
\begin{array}{cc}
I & 0 \\ 
0 & -I%
\end{array}%
\right) \left( 
\begin{array}{c}
T_{1} \\ 
T_{2}%
\end{array}%
\right) >0\quad \Longleftrightarrow \quad I-z^{\dagger }z>0  \label{4c1}
\end{equation}%
is the bounded\textbf{\ }$SU(2,2)$ classical domain\cite{PIAT1966}, because
it is bounded in the general $z$-space and it is invariant under the $%
SU(2,2) $\textbf{\ }transformation%
\begin{equation}
\begin{array}{l}
\begin{pmatrix}
T_{1}^{\prime } \\ 
T_{2}^{\prime }%
\end{pmatrix}%
=%
\begin{pmatrix}
A_{11} & A_{12} \\ 
A_{21} & A_{22}%
\end{pmatrix}%
\left( 
\begin{array}{c}
T_{1} \\ 
T_{2}%
\end{array}%
\right) \\ 
\\ 
z^{\prime }=\left( A_{21}+A_{22}\ z\right) \left( A_{11}+A_{12}\ z\right)
^{-1} \\ 
\\ 
A_{11}^{\dagger }A_{11}-A_{21}^{\dagger }A_{21}=I\quad ,\quad
A_{11}^{\dagger }A_{12}-A_{21}^{\dagger }A_{22}=0\quad ,\quad
A_{22}^{\dagger }A_{22}-A_{12}^{\dagger }A_{12}=I \\ 
\end{array}
\label{4c2}
\end{equation}

The characteristic (Shilov) boundary of this domain is the $S^{1}\times
S^{3}[=U(2)]$ manifold with $z^{\dagger }z=I$. The ordinary parametrization
of this boundary is 
\begin{equation}
\begin{array}{l}
U=e^{i\tau }\left( 
\begin{array}{cc}
\cos \rho +i\sin \rho \cos \theta & i\sin \rho \sin \theta \ e^{-i\varphi }
\\ 
i\sin \rho \sin \theta \ e^{i\varphi } & \cos \rho -i\sin \rho \cos \theta%
\end{array}%
\right) = \\ 
\\ 
\qquad =\frac{1+r^{2}-t^{2}+2it}{[1+2(t^{2}+r^{2})+(t^{2}-r^{2})^{2}]}\left( 
\begin{array}{cc}
1+t^{2}-r^{2}-2iz & -2i(x-iy) \\ 
-2i(x+iy) & 1+t^{2}-r^{2}+2iz%
\end{array}%
\right)%
\end{array}
\label{4c3}
\end{equation}%
where $\tau \in (-\pi ,\pi )\ ,\ \varphi \in (0,2\pi )\ ,\ \rho \in (0,\pi
)\ ,\ \theta \in (0,\pi )$.

In the homogeneous coordinates%
\begin{equation}
\begin{array}{l}
H=\left( 
\begin{array}{c}
H_{1} \\ 
H_{2}%
\end{array}%
\right) =\frac{1}{\sqrt{2}}\left( 
\begin{array}{cc}
I & -I \\ 
I & I%
\end{array}%
\right) \left( 
\begin{array}{c}
T_{1} \\ 
T_{2}%
\end{array}%
\right) \\ 
\\ 
T=\left( 
\begin{array}{c}
T_{1} \\ 
T_{2}%
\end{array}%
\right) =\frac{1}{\sqrt{2}}\left( 
\begin{array}{cc}
I & I \\ 
-I & I%
\end{array}%
\right) \left( 
\begin{array}{c}
H_{1} \\ 
H_{2}%
\end{array}%
\right)%
\end{array}
\label{4d1}
\end{equation}%
we have%
\begin{equation}
\left( 
\begin{array}{cc}
0 & I \\ 
I & 0%
\end{array}%
\right) =\frac{1}{2}\left( 
\begin{array}{cc}
I & -I \\ 
I & I%
\end{array}%
\right) \left( 
\begin{array}{cc}
I & 0 \\ 
0 & -I%
\end{array}%
\right) \left( 
\begin{array}{cc}
I & I \\ 
-I & I%
\end{array}%
\right)  \label{4d2}
\end{equation}%
and the positive definite condition takes the form%
\begin{equation}
\begin{pmatrix}
H_{1}^{\dagger } & H_{2}^{\dagger }%
\end{pmatrix}%
\left( 
\begin{array}{cc}
0 & I \\ 
I & 0%
\end{array}%
\right) \left( 
\begin{array}{c}
H_{1} \\ 
H_{2}%
\end{array}%
\right) >0\quad \Longleftrightarrow \quad -i(r-r^{\dagger })=y>0  \label{4d3}
\end{equation}%
where the projective coordinates $r_{A^{\prime }B}=x_{A^{\prime
}B}+iy_{A^{\prime }B}$ are defined as $r=iH_{2}H_{1}^{-1}$, which implies $%
H_{2}=-irH_{1}$ and%
\begin{equation}
\begin{array}{l}
r=i(I+z)(I-z)^{-1}=i(I-z)^{-1}(I+z) \\ 
\\ 
z=(r-iI)(r+iI)^{-1}=(r+iI)^{-1}(r-iI)%
\end{array}
\label{4d4}
\end{equation}%
The fractional transformations which preserve the unbounded domain are%
\begin{equation}
\begin{array}{l}
\begin{pmatrix}
H_{1}^{\prime } \\ 
H_{2}^{\prime }%
\end{pmatrix}%
=%
\begin{pmatrix}
B_{11} & B_{12} \\ 
B_{21} & B_{22}%
\end{pmatrix}%
\left( 
\begin{array}{c}
H_{1} \\ 
H_{2}%
\end{array}%
\right) \\ 
\\ 
r^{\prime }=\left( B_{22}\ r+iB_{21}\right) \left( B_{11}-iB_{12}\ r\right)
^{-1} \\ 
\\ 
B_{11}^{\dagger }B_{22}+B_{21}^{\dagger }B_{12}=I\quad ,\quad
B_{11}^{\dagger }B_{21}+B_{21}^{\dagger }B_{11}=0\quad ,\quad
B_{22}^{\dagger }B_{12}+B_{12}^{\dagger }B_{22}=0%
\end{array}
\label{4d5}
\end{equation}%
The characteristic boundary in this "upper plane" realization of the
classical domain is the "real axis"%
\begin{equation}
y_{A^{\prime }A}=0  \label{4d6}
\end{equation}

Using the $SU(2,2)$ generators, the $SL(4,C)$ infinitesimal transformations
have the form%
\begin{equation}
\begin{tabular}{l}
$\delta Y=\frac{i}{2}\epsilon ^{\mu }\gamma _{\mu }(1+\gamma _{5})Y\quad
,\quad \delta Y=-\frac{i}{4}\epsilon ^{\mu \nu }\sigma _{\mu \nu }Y$ \\ 
\\ 
$\delta Y=-\frac{i}{2}\kappa ^{\mu }\gamma _{\mu }(1-\gamma _{5})Y\quad
,\quad \delta Y=-\frac{1}{2}\rho \gamma _{5}Y$ \\ 
\end{tabular}
\label{b13}
\end{equation}%
The real parts of the infinitesimal variables ($\epsilon ^{\mu }$, $\epsilon
^{\mu \nu }$,\ $\kappa ^{\mu }$,\ $\rho $) provide the 15 $SU(2,2)$ charges
and their imaginary parts the remaining 15 charges of the $SL(4,C)$
transformations.

Considering the explicit forms of the homogeneous coordinates we have%
\begin{equation}
H=%
\begin{pmatrix}
X^{01} & X^{02} \\ 
X^{11} & X^{12} \\ 
X^{21} & X^{22} \\ 
X^{31} & X^{32}%
\end{pmatrix}%
=%
\begin{pmatrix}
\lambda ^{01} & \lambda ^{02} \\ 
\lambda ^{11} & \lambda ^{12} \\ 
-i(r_{0^{\prime }0}\lambda ^{01}+r_{0^{\prime }1}\lambda ^{11}) & 
-i(r_{0^{\prime }0}\lambda ^{02}+r_{0^{\prime }1}\lambda ^{12}) \\ 
-i(r_{1^{\prime }0}\lambda ^{01}+r_{1^{\prime }1}\lambda ^{11}) & 
-i(r_{1^{\prime }0}\lambda ^{02}+r_{1^{\prime }1}\lambda ^{12})%
\end{pmatrix}
\label{4d7}
\end{equation}%
where everything has been arranged such that the spinor transformations
imply the corresponding spacetime transformations and vice-versa.

If we restrict the above $z_{ij}\rightarrow r_{A^{\prime }B}$\
transformations at the Shilov boundary we find the form%
\begin{equation}
\begin{array}{l}
t=\frac{\sin \tau }{\cos \tau \ -\ \cos \rho } \\ 
\\ 
x+iy=\frac{\sin \rho }{\cos \tau \ -\cos \rho }\sin \theta \ e^{i\varphi }
\\ 
\\ 
z=\frac{\sin \rho }{\cos \tau \ -\ \cos \rho }\cos \theta%
\end{array}
\label{4d9}
\end{equation}%
Additional formulas are%
\begin{equation}
\begin{array}{l}
r=\frac{\sin \rho }{\cos \tau \ -\ \cos \rho }=\frac{-\sin \rho }{2\sin 
\frac{\tau +\rho }{2}\sin \frac{\tau -\rho }{2}} \\ 
\\ 
\sqrt{1+2(t^{2}+r^{2})+(t^{2}-r^{2})^{2}}=\frac{2}{\cos \tau \ -\cos \rho }%
\end{array}
\label{4d10}
\end{equation}%
where now $r=\sqrt{x^{2}+y^{2}+z^{2}}$ denotes the ordinary radial component
and takes positive values. Notice that the Cartesian coordinates $(x,y,z)$
are the projective coordinates from the center of $S^{3}$. We essentially
need two such tangent planes to cover the whole sphere (but equator). The
two hemispheres are covered by permitting the radial variable $r$ to take
negative values too.

We also see that%
\begin{equation}
\begin{array}{l}
t-r=-\cot \frac{\tau -\rho }{2} \\ 
\\ 
t+r=-\cot \frac{\tau +\rho }{2}%
\end{array}
\label{4d11}
\end{equation}

Through the above transformation Minkowski spacetime is conformally
equivalent to the half of\textbf{\ }$S^{1}\times S^{3}$. It is easy to prove
that the following two points of $S^{1}\times S^{3}$ correspond to the same
point $(t,x,y,z)$ of the Minkowski space. 
\begin{equation}
\begin{array}{l}
(\tau ,\rho ,\theta ,\varphi )\Longrightarrow (t,x,y,z) \\ 
\\ 
(\tau +\pi \ ,\ \pi -\rho \ ,\ \pi -\theta \ ,\ \varphi +\pi
)\Longrightarrow (t,x,y,z)%
\end{array}
\label{4d12}
\end{equation}%
In the $\tau ,\rho $ axes\ the $r=\infty $ boundaries are $\tau -\rho =0$
and $\tau +\rho =0$. These are the two Penrose boundaries $\mathfrak{J}^{\pm
}$ of Minkowski spacetime. The two triangles at both sides of the $\rho $
axis have $r>0$ and the other two triangles have $r<0$.

Minkowski spacetime may be \textquotedblleft properly" compactified by
simply identifying its two Penrose boundaries $\mathfrak{J}^{+}$ and $%
\mathfrak{J}^{-}$. This is naturally done through the identification of
compactified Minkowski spacetime with the characteristic boundary of the
type IV $SO(2,4)$\ invariant classical domain, which is the part of $CP^{5}$
determined by the relations\cite{PIAT1966}%
\begin{equation}
t^{\intercal }Ht=0\quad ,\quad t^{\dagger }Ht>0\quad ,\quad \func{Im}\frac{%
t_{1}}{t_{0}}>0  \label{4d13}
\end{equation}%
where $t$ is\ a 6-dimensional complex column (the homogeneous coordinates of 
$CP^{5}$) and $H=diag[1,1,-1,-1,-1,-1]$. Because of the homomorphism between 
$SU(2,2)$ and $O(2,4)$,\ the homogeneous coordinates ($X^{mi}$) of $G_{2,2}$
are related to the homogeneous coordinates of $CP^{5}$ with the following
relations\cite{P-R1984}\ 
\begin{equation}
\begin{array}{l}
t_{0}=\frac{i}{\sqrt{2}}(R^{12}-R^{03})\quad ,\quad t_{1}=R^{01}+\frac{1}{2}%
R^{23} \\ 
\\ 
t_{2}=R^{01}-\frac{1}{2}R^{23}\quad ,\quad t_{3}=\frac{i}{\sqrt{2}}%
(R^{02}-R^{13}) \\ 
\\ 
t_{4}=\frac{1}{\sqrt{2}}(R^{02}+R^{13})\quad ,\quad t_{5}=\frac{-i}{\sqrt{2}}%
(R^{12}+R^{03}) \\ 
\end{array}
\label{4d14}
\end{equation}%
where $R^{mn}=X^{m1}X^{n2}-X^{n1}X^{m2}$. Recall that the homomorphisms
between the three conformal groups are%
\begin{equation}
SU(2,2)\overset{2\rightarrow 1}{===>}O_{+}^{\uparrow }(2,4)\overset{%
2\rightarrow 1}{===>}C_{+}^{\uparrow }(1,3)  \label{4d15}
\end{equation}

\subsection{Complex structures in $G_{2,2}$ context}

In the simple case of conformally flat spacetimes, the integrability
condition of the complex structure can be solved by Kerr's theorem\cite%
{FLAHE1974}. Using the $G_{2,2}$ homogeneous coordinates $X^{mi}$\ this
general solution takes the form\cite{P-R1984}%
\begin{equation}
\begin{array}{l}
\overline{X^{mi}}E_{mn}X^{nj}=0 \\ 
\\ 
K_{i}(X^{mi})=0%
\end{array}
\label{4e1}
\end{equation}%
where the first line relations fix the surface to the Shilov boundary or a
part of it, and the second line relations are the two Kerr homogeneous
function. The structure coordinates $z^{\alpha }$ are then two independent
functions of $\frac{X^{m1}}{X^{01}}$ and $z^{\widetilde{\alpha }}$ are two
independent functions of $\frac{X^{m2}}{X^{02}}$.

This particular solution indicates the form of a general solution for the
integrability condition of the complex structure on a generally curved
spacetime. In this case the $G_{2,2}$ homogeneous coordinates $X^{mi}$\ have
to satisfy relations of the form%
\begin{equation}
\begin{array}{l}
\Omega _{ij}(\overline{X^{mi}},X^{nj})=0 \\ 
\\ 
K_{i}(X^{mi})=0%
\end{array}
\label{4e2}
\end{equation}%
where all the functions are homogeneous relative to $X^{n1}$ and $X^{n2}$\
independently. That is, they are defined in $CP^{3}\times CP^{3}$. The
rank-2 condition on the matrix $X^{mi}$ defines the solutions as $SL(2,C)$
fiber bundles on 4-dimensional surfaces of $G_{2,2}$. The structure
coordinates $(z^{\alpha }\ ,\ z^{\widetilde{\alpha }})$ are determined
exactly like in the simple case of conformally flat spacetimes given above.

In General Relativity the asymptotic flatness condition is imposed using the
metric. In the present case of complex structures this condition is imposed
through the assumption that there are independent homogeneous
transformations of $X^{n1}$ and $X^{n2}$\ such that 
\begin{equation}
\begin{array}{l}
\overline{X^{m1}}E_{mn}X^{n1}=0 \\ 
\overline{X^{m2}}E_{mn}X^{n2}=0 \\ 
\overline{X^{m1}}E_{mn}X^{n2}\neq 0 \\ 
\end{array}
\label{4e3}
\end{equation}%
That is the two functions $\Omega _{11}(\overline{X^{m1}},X^{n1})$ and\ $%
\Omega _{22}(\overline{X^{m2}},X^{n2})$\ take the flat space forms.\ The
first two annihilations will be used below to restrict the forms of
stationary axisymmetric complex structures.

The central problem of the present work is to find solutions $X^{mi}(x)$\
which satisfy relations of the form (\ref{4e2}). The topological classes of
these solutions will determine the soliton sectors of the model. The
algebraic nature of the two homogeneous Kerr functions $K_{i}(X^{mi})$\ is a
powerful mathematical property which will be used below. But from the
physical point of view it is somehow obscure because it hides physical
intuition. Therefore one way to replace them is the parametrization (\ref%
{4a1}) of $X^{mi}$\ where $\lambda ^{Ai}$ are functions of $r^{A^{\prime }A}$
which satisfy the Kerr differential equations 
\begin{equation}
\begin{array}{l}
\lambda ^{A}\lambda ^{B}\frac{\partial }{\partial r^{A^{\prime }A}}\lambda
_{B}=0 \\ 
\end{array}
\label{4e4}
\end{equation}%
which was our intuitive procedure for the discovery of the form (\ref{4e2})
of the general solution.

Another physically very intuitive form, which replaces the Kerr functions,
is the following trajectory parametrization of $X^{mi}$\ 
\begin{equation}
X^{mi}=%
\begin{pmatrix}
\lambda ^{Ai} \\ 
-i\xi _{A^{\prime }B}^{i}(\tau _{i})\lambda ^{Bi}%
\end{pmatrix}
\label{4e5}
\end{equation}%
where $\xi _{A^{\prime }B}^{i}(\tau _{i})$\ are two complex trajectories in
the Grassmannian manifold $G_{2,2}$.\ A combination of this parametrization
with the Grassmannian one (\ref{4a1}) implies the two conditions $\det
[r_{A^{\prime }B}-\xi _{A^{\prime }B}^{i}(\tau _{i})]=0$ for the two linear
equations $[r_{A^{\prime }B}-\xi _{A^{\prime }B}^{i}(\tau _{i})]\lambda
^{Bi}=0$\ to admit non-vanishing solutions. Notice that this condition
(restricted to the Shilov boundary) is identical to the relation 
\begin{equation}
\eta _{\mu \nu }(x^{\mu }-\xi ^{\mu }(\tau ))(x^{\nu }-\xi ^{\nu }(\tau
))=0\;  \label{4ee6}
\end{equation}%
which was first used by Newman and coworkers\cite{N-W1974} to determine
twisted geodetic and shear free null congruences in Minkowski spacetime.

It is to prove that the quadratic Kerr polynomial%
\begin{equation}
Z^{1}Z^{2}-Z^{0}Z^{3}+2aZ^{0}Z^{1}=0  \label{4ee7}
\end{equation}%
is implied by the trajectory $\xi ^{\mu }(\tau )=(\tau \ ,\ 0\ ,\ 0\ ,\ ia)$.

In the case of one trajectory, we will have one Kerr function. In this case
and for a trajectory normalized by $\xi ^{0}(\tau )=\tau $\ , the asymptotic
flatness condition implies%
\begin{equation}
i(\overline{\tau }-\tau )-i(\overline{\xi ^{1}}-\xi ^{1})\frac{\lambda +%
\overline{\lambda }}{1+\lambda \overline{\lambda }}+(\overline{\xi ^{2}}-\xi
^{2})\frac{\overline{\lambda }-\lambda }{1+\lambda \overline{\lambda }}-i(%
\overline{\xi ^{3}}-\xi ^{3})\frac{1+\lambda \overline{\lambda }}{1+\lambda 
\overline{\lambda }}=0  \label{4e6}
\end{equation}%
which fixes the imaginary parts of the two complex parameters $\tau _{1}$
and $\tau _{2}$.

Using the $G_{2,2}$ formalism, the (\ref{2d3}) complex structure of
Minkowski spacetime is implied by the quadratic homogeneous polynomial (\ref%
{4ee7}). Notice that the points of spacetime (\ref{2d8}) with $\det (\lambda
^{Ai})=0$\ do not belong to the Grassmannian projective manifold, because
the corresponding $4\times 2$ matrix has not rank 2. This means that if the
spacetime is defined as a surface of of $G_{2,2}$, these points do not
belong to the spacetime!

Using the Kerr-Schild ansatz, we have derived a general curved complex
structure (\ref{2e2}). Assuming that the Kerr function of the curved complex
structure is the same with that of the corresponding flatprint complex
structure we can find the 4-dimensional surface of $G_{2,2}$. Its explicit
form for the Kerr-Newman complex structure ($h(r)=-2mr+q^{2}$) in Lindquist
coordinates $(t,r,\theta ,\varphi )$ is%
\begin{equation}
\begin{array}{l}
x^{0}=t+\frac{f_{1}}{\cos ^{2}f+\cos ^{2}\theta \sin ^{2}f} \\ 
\\ 
x^{1}+ix^{2}=\frac{(r+f_{1})\cos f+a\sin ^{2}\theta \sin f}{\cos ^{2}f+\cos
^{2}\theta \sin ^{2}f}\sin \theta e^{i\varphi }e^{-if} \\ 
\\ 
x^{3}=r\cos \theta +\frac{f_{1}\cos \theta }{\cos ^{2}f+\cos ^{2}\theta \sin
^{2}f}+ \\ 
\qquad +\frac{\sin f\sin ^{2}\theta }{\cos ^{2}f+\cos ^{2}\theta \sin ^{2}f}%
[(r\sin f-a\cos f)\cos \theta +f_{1}\sin f] \\ 
\\ 
y^{0}=0 \\ 
\\ 
y^{1}+iy^{2}=\frac{(r+f_{1})\sin f-a\cos f}{\cos ^{2}f+\cos ^{2}\theta \sin
^{2}f}\cos \theta \sin \theta e^{i\varphi }e^{-if} \\ 
\\ 
y^{3}=\frac{\cos f\sin ^{2}\theta }{\cos ^{2}f+\cos ^{2}\theta \sin ^{2}f}%
[a\cos f-(r+f_{1})\sin f]%
\end{array}%
\end{equation}%
where the two functions entering the configuration are

\begin{equation}
\begin{array}{l}
f=\frac{a}{2\sqrt{a^{2}+q^{2}-m^{2}}}\arctan \frac{2(r-m)\sqrt{%
a^{2}+q^{2}-m^{2}}}{r^{2}-2mr++2m^{2}-a^{2}-q^{2}} \\ 
\\ 
f_{1}=-m\ln \frac{r^{2}-2mr+a^{2}+q^{2}}{m^{2}}+\frac{2m^{2}-q^{2}}{a}f%
\end{array}
\label{5h14}
\end{equation}%
Notice that this surface is outside the classical domain because $y^{0}=0$.

As a second example we will present below the \textquotedblleft
natural\textquotedblright\ complex structure of the $U(2)$ surface. The
1-forms of the left invariant generators of the group $U(2)$ are defined by
the relation $U^{\dag }dU=ie_{L}^{a}\sigma _{a}=i(e_{L}^{0}\sigma
^{0}-e_{L}^{i}\sigma ^{i})$. In the (\ref{4c3}) parametrization the 1-forms
are%
\begin{equation}
\begin{array}{l}
e_{L}^{0}=d\tau \\ 
\\ 
e_{L}^{1}=-\sin \theta \cos \varphi d\rho +(\sin ^{2}\rho \sin \varphi -\sin
\rho \cos \rho \cos \theta \cos \varphi )d\theta + \\ 
\qquad +(\sin ^{2}\rho \cos \theta \sin \theta \cos \varphi +\sin \rho \cos
\rho \sin \theta \sin \varphi )d\varphi \\ 
\\ 
e_{L}^{2}=-\sin \theta \sin \varphi d\rho +(-\sin \rho \cos \rho \cos \theta
\sin \varphi +\sin ^{2}\rho \cos \varphi )d\theta + \\ 
\qquad +(\sin ^{2}\rho \cos \theta \sin \theta \sin \varphi -\sin \rho \cos
\rho \sin \theta \cos \varphi )d\varphi \\ 
\\ 
e_{L}^{3}=-\cos \theta d\rho +\sin \rho \cos \rho \sin \theta d\theta -\sin
^{2}\rho \sin ^{2}\theta d\varphi%
\end{array}
\label{4k2}
\end{equation}

In Cartesian coordinates the generators take the form [$C=\frac{4}{%
1+2(t^{2}+r^{2})+(t^{2}-r^{2})^{2}}$]%
\begin{equation}
\begin{array}{l}
e_{L}^{0}=C[(1+r^{2}+t^{2})dt-2txdx-2tydy-2tzdz] \\ 
\\ 
e_{L}^{1}=C[-2xtdt+(1+t^{2}+x^{2}-y^{2}-z^{2})dx+ \\ 
\qquad +2(xy+z)dy+2(xz-y)dz] \\ 
\\ 
e_{L}^{2}=C[-2ytdt+2(xy-z)dx+(1+t^{2}-x^{2}+y^{2}-z^{2})dy+ \\ 
\qquad +2(yz+x)dz] \\ 
\\ 
e_{L}^{3}=C[-2ztdt+2(xz+y)dx+2(yz-x)dy+ \\ 
\qquad +(1+t^{2}-x^{2}-y^{2}+z^{2})dz]%
\end{array}
\label{4k4}
\end{equation}

The 1-forms satisfy the following differential relations%
\begin{equation}
de_{L}^{0}=0\quad ,\quad de_{L}^{i}=\epsilon _{ijk}e_{L}^{j}\wedge e_{L}^{k}
\label{4k8}
\end{equation}%
which imply the relations%
\begin{equation}
\left( e^{i}e^{j}\partial e^{k}\right) =e^{i\mu }e^{j\nu }\left( \partial
_{\mu }e_{\nu }^{k}-\partial _{\nu }e_{\mu }^{k}\right) =2\epsilon _{ijk}
\label{4k9}
\end{equation}

The \textquotedblleft natural\textquotedblright\ complex structure on $%
S^{1}\times S^{3}$\ is defined by the following tetrad%
\begin{equation}
\begin{array}{l}
L^{\mu }=e_{L}^{0\mu }-e_{L}^{3\mu } \\ 
\\ 
N^{\mu }=e_{L}^{0\mu }-e_{L}^{3\mu } \\ 
\\ 
M^{\mu }=e_{L}^{1\mu }+ie_{L}^{2\mu }%
\end{array}
\label{4l1}
\end{equation}

This complex structure is generated from the following degenerate quadratic
polynomial, which is the product of two linear polynomials%
\begin{equation}
(Z^{1}+Z^{3})(Z^{0}+Z^{2})=0  \label{4l2}
\end{equation}

The surface is the boundary of the classical domain $y_{A^{\prime }A}=0$
with the following homogeneous coordinates of $G_{2,2}$ 
\begin{equation}
H=%
\begin{pmatrix}
1 & \frac{x-iy}{t-z+i} \\ 
\frac{x+iy}{t+z+i} & 1 \\ 
-i\left[ t-z-(x-iy)\lambda _{1}\right] & \frac{-i}{\lambda _{2}}\left[
t-z-(x-iy)\lambda _{2}\right] \\ 
-i\left[ -(x+iy)+(t+z)\lambda _{1}\right] & \frac{-i}{\lambda _{2}}\left[
-(x+iy)+(t+z)\lambda _{2}\right]%
\end{pmatrix}
\label{4l3}
\end{equation}%
These structure coordinates are valid over the whole Shilov boundary space
because $\lambda _{1}\neq \lambda _{2}$ everywhere, while the corresponding
surface of $CP^{3}$, defined by the Kerr polynomial (\ref{4l2}) is singular.

\subsection{Induced metrics on spacetimes}

The characteristic property of the present model is that it depends on the
integrable complex structure $J_{\mu }^{\ \ \nu }$ and not on a metric $%
g_{\mu \nu }$. The above approach to the solution of the complex structure
equations through the $G_{2,2}$ mathematical machinery permit us to look at
spacetimes from a different point of view. The metric should be seen as a
product and not as a \textquotedblleft primitive\textquotedblright\
dynamical variable. In fact we may define more than one metric on the
4-dimensional surface of $G_{2,2}$. The complex structure determines its
eigenvectors $(\ell _{\mu },n_{\mu },m_{\mu },\overline{m}_{\mu })$ up to
four independent Weyl factors. These vectors permit us to define a symmetric
tensor $g_{\mu \nu }$ through formula (\ref{2a4}). This tensor may be used
as a metric of the spacetime. The tetrad is apparently null relative to this
metric, which is compatible with the complex structure $J_{\mu }^{\ \ \nu }$%
. In fact the most general metric we may use is 
\begin{equation}
g_{\mu \nu }=\Omega ^{2}\left[ (\ell _{\mu }n_{\nu }+n_{\mu }\ell _{\nu
})-\omega ^{2}\ (m_{{}\mu }\overline{m}_{\nu }+\overline{m}_{\mu }m_{\nu })%
\right]  \label{4p1}
\end{equation}%
where $\Omega (x)$ and $\omega (x)$ are two arbitrary real functions. I want
to point out that this metric arbitrariness saves the present model from the
scalar solitons which caused the most serious difficulties to the
Misner-Wheeler geometrodynamic model. Notice that the spherically symmetric
spacetimes (e.g. Schwarzschild) are compatible with the Minkowski metric.
Therefore they do not differ from the \textquotedblleft
vacuum\textquotedblright\ surface.

The above metric is very useful because it is directly related to the
complex structure. In fact it determines the complex structure through the
algebraic condition (\ref{2a13}), and in the \textquotedblleft
vacuum\textquotedblright\ surface (Minkowski spacetime) it may take a form
which respects the remaining Poincar\'{e} symmetry. But it is not the only
metric we may define. The rank-2 matrices $X^{mi}(\xi )$ permit us to induce
the well known $SU(4)$ and $SU(2,2)$ invariant metrics of $G_{2,2}$ down to
the 4-dimensional surface. In the bounded (Dirac representation) coordinate
neighborhood the surface is%
\begin{equation}
\begin{array}{l}
z_{11}=\frac{X^{21}X^{12}-X^{11}X^{22}}{X^{01}X^{12}-X^{11}X^{02}}\quad
,\quad z_{12}=\frac{X^{01}X^{22}-X^{21}X^{02}}{X^{01}X^{12}-X^{11}X^{02}} \\ 
\\ 
z_{21}=\frac{X^{31}X^{12}-X^{11}X^{32}}{X^{01}X^{12}-X^{11}X^{02}}\quad
,\quad z_{22}=\frac{X^{01}X^{32}-X^{31}X^{02}}{X^{01}X^{12}-X^{11}X^{02}}%
\end{array}
\label{4p3}
\end{equation}

In this coordinate neighborhood the $SU(4)$ and $SU(2,2)$ invariant metrics
of $G_{2,2}$ are%
\begin{equation}
ds_{\pm }^{2}=\frac{\partial }{\partial \overline{z_{ij}}}\frac{\partial }{%
\partial z_{kl}}\ln [\det (I\pm z^{\dag }z)]\ d\overline{z_{ij}}\ dz_{kl}
\label{4p4}
\end{equation}%
where the \textquotedblleft +\textquotedblright\ and \textquotedblleft
-\textquotedblright\ denote the $SU(4)$ and $SU(2,2)$ invariant metrics
respectively. These metrics may easily be transcribed in the unbounded
(chiral) coordinate neighborhood. After the direct substitution of $%
z_{ij}(\xi )$, the induced Euclidean metrics on the 4-dimensional surface
may be found. These metrics do not seem to be directly related to the
complex structure of the surface.\newpage

\section{VACUA\ AND\ EXCITATION\ MODES}

\setcounter{equation}{0}

A physically interesting geometrodynamic model must generate the
electromagnetic field and the intermediate vector bosons of the Standard
Model from the fundamental equations of the model itself, without
introducing anything by hand. In the Misner-Wheeler model, the
electromagnetic potential is directly and exactly derived from the Rainich
conditions. In fact this derivation was the essential reason behind the
assumption of the Rainich conditions, as the fundamental equations of the
Misner-Wheeler model\cite{M-W1957}. In Quantum Field Theory the vacuum
excitation modes are the periodic configurations which diagonalize energy
and momentum. In the present context periodicity is understood in
compactified Minkowski spacetime \textrm{M}$^{\mathrm{\#}}$. Therefore we
first have to define energy and momentum, and after to look for the model
vacua with vanishing energy and momentum, the excitation modes and the
solitons with finite energy.

\subsection{Physical energy-momentum}

It is well known that in any generally covariant model the translation
generators are first class constraints, which must vanish. Therefore energy,
momentum and angular momentum cannot be defined using Noether's theorem. The
success of the Einstein equations strongly suggests that energy-momentum has
to be defined through the Einstein tensor $E^{\mu \nu }$. The direct
relation of the Einstein tensor with the classical energy-momentum and
angular momentum is also strongly implied by the derivation of the equations
of motion in the harmonic coordinate system, imposed by the condition $%
\partial _{\mu }\left( \sqrt{-g}g^{\mu \nu }\right) =0$, using the
contracted Bianchi identities $\nabla _{\mu }E^{\mu \nu }=0$. But $E^{\mu
\nu }$\ depends on the metric and it is not directly related to the Poincar%
\'{e} generators of the present model. Therefore for the definition of the
energy in the present model we proceed as follows\cite{RAG1999}.

We first consider the coordinate system imposed by the relation 
\begin{equation}
\partial _{\mu }\left( \sqrt{-g}E^{\mu \nu }\right) =0  \label{6g1}
\end{equation}%
We next consider the conserved quantity 
\begin{equation}
\mathcal{E}(g^{\mu \nu })=\int_{\!t}\sqrt{-g}E^{\mu 0}dS_{\mu }  \label{6g3}
\end{equation}%
where the time variable $t$ is chosen such that $\mathcal{E}(g^{\mu \nu
})\geq 0$. This quantity depends on the metric $g^{\mu \nu }$ and it does
not characterize the complex structure, therefore it cannot be the energy
definition of the configuration. We think that the energy of a complex
structure is properly defined by the following minimum 
\begin{equation}
\mathrm{E}[J_{\mu }^{\;\nu }]=\underset{g_{\mu \nu }\in \lbrack J_{\mu
}^{\;\nu }]}{\min }\ \mathcal{E}(g^{\mu \nu })  \label{6g4}
\end{equation}%
where the minimum is taken over all the class $[J_{\mu }^{\;\nu }]$ of
metrics (\ref{4p1}).

Apparently this conserved quantity depends only on the moduli parameters of
the complex structure. In a vacuum sector it vanishes, $\mathrm{E}[J_{\mu
}^{\;\nu }]=0$. From the 2-dimensional solitonic models\cite{FELS1981}, we
know that the minima of the energy characterize the solitons. Assuming that $%
\mathrm{E}[J_{\mu }^{\;\nu }]$ is a smooth function of the moduli
parameters, we can always expand it around a minimum. 
\begin{equation}
\mathrm{E}[J_{\mu }^{\;\nu }]\simeq E+\underset{q}{\sum }\varepsilon _{q}\ 
\overline{a_{q}}\ a_{q}  \label{6g6}
\end{equation}%
where $E$ and $\varepsilon _{q}$ are positive parameters. These variables
and $a_{q}$ are moduli parameters of the complex structure. $E$ is defined
to be the energy of the soliton characterized by the minimum and $%
\varepsilon _{q}$ are the energies of the excitation modes. In the special
metric where the minimum (\ref{6g4}) occurs, we can define the 4-momentum
and the angular momentum 
\begin{equation}
\begin{array}{l}
\mathrm{P}^{\nu }=\int_{\!t}\sqrt{-g}E^{\mu \nu }dS_{\mu } \\ 
\\ 
\mathrm{S}^{\mu \nu }=\int_{\!t}\sqrt{-g}E^{\rho \sigma }x^{\tau }\ \Sigma
_{\sigma \tau }^{\mu \nu }\ dS_{\mu }%
\end{array}
\label{6g7}
\end{equation}

These quantities are conserved in the precise coordinate systems, which
satisfy (\ref{6g1}). But this is not enough to identify them with the Poincar%
\'{e} group generators! Recall that the Poincar\'{e} transformations are
well defined in the present model. They form a subalgebra of $sl(4,C)$ and a
part of the infinite algebra of the complex structure preserving
transformations. The relation of these Poincar\'{e} group generators with
the present conserved quantities is implied by the transformation of the
Einstein tensor under the Poincar\'{e} transformations. Recall that in the
unbounded coordinate neighborhood of $G_{2,2}$ the Poincar\'{e}
transformations do not mix the Hermitian $x_{A^{\prime }A}$ and the
anti-Hermitian $iy_{A^{\prime }A}$ parts of the projective coordinates $%
r_{A^{\prime }A}$. Therefore we must first fix the coordinates to be the
Cartesian coordinate system defined as the real part of the $r^{A^{\prime
}A} $ projective coordinates of the Grassmannian manifold $G_{2,2}$. But in
this coordinate system energy-momentum is not exactly conserved. It is
approximately conserved in the \textquotedblleft weak
gravity\textquotedblright\ limit. Therefore we will consider the modes which
diagonalize this \textquotedblleft weak gravity\textquotedblright\ limit of
energy-momentum. These modes belong to irreducible representations of the
Poincar\'{e} transformations, properly defined on the Cartesian coordinate
system as $\delta x^{\mu }=\omega _{\quad \nu }^{\mu }x^{\nu }+\varepsilon
^{\mu }$. Then the Quantum Theory relation 
\begin{equation}
i[\varepsilon Q_{\varepsilon }\ ,\ E^{\mu \nu }]=\delta _{\varepsilon
}E^{\mu \nu }=E^{\mu \rho }\partial _{\rho }\varepsilon ^{\nu }+E^{\rho \nu
}\partial _{\rho }\varepsilon ^{\mu }-\varepsilon ^{\rho }\partial _{\rho
}E^{\mu \nu }  \label{6g8}
\end{equation}%
implies that $\mathrm{P}^{\nu }$ approximately behaves as a vector and $%
\mathrm{S}^{\mu \nu }$ as an antisymmetric tensor. $\mathrm{P}^{\mu }$\ and $%
\mathrm{S}_{z}$\ commute with the corresponding Poincar\'{e} generators.
Hence the approximative relation (\ref{6g6}) and the preceding Poincar\'{e}
group transformations imply the forms 
\begin{equation}
\begin{array}{c}
\mathrm{P}^{\mu }\simeq k_{0}^{\mu }+\underset{i,s}{\sum }\int {}d^{3}k\
k^{\mu }\ a_{i}^{+}(\overset{\rightarrow }{k},s)\ a_{i}(\overset{\rightarrow 
}{k},s) \\ 
\\ 
\mathrm{S}_{z}\simeq s_{z}+\underset{i,s}{\sum }\int {}d^{3}k\ s\ a_{i}^{+}(%
\overset{\rightarrow }{k},s)\ a_{i}(\overset{\rightarrow }{k},s)%
\end{array}
\label{6g9}
\end{equation}%
where the summation is over the momentum, the spin and the irreducible
representations $i$ of the Poincar\'{e} group. $k_{0}^{\mu }$ is the
4-momentum and $s_{z}$ is the z-component of the spin of the soliton. $%
k^{\mu }$ is the 4-momentum and $s$ is the z-component of the spin of the
excitation modes $a_{i}(\overset{\rightarrow }{k},s)$. In the quantized
theory the variables $a_{i}^{+}(\overset{\rightarrow }{k},s)$ and $a_{i}(%
\overset{\rightarrow }{k},s)$ become the creation and the annihilation
operators of the approximative modes, which diagonalize the 4-momentum. From
Quantum Field Theory we know that the second parts of (\ref{6g9}) are
formally generated by the ordinary energy momentum tensors of free quantum
fields. They should be bosonic because they represent excitation modes. This
procedure permit us to write down the Einstein tensor as the energy-momentum
tensor of the excitation modes. Notice that this effective energy-momentum
tensor has to contain interactions, because the field excitation modes
diagonalize the approximative Einstein tensor. We will refer to this
effective Lagrangian again below in relation to the computation of the
soliton form factors.

The tetrad vectors $(\ell _{\mu },\,n_{\mu },\,m_{\mu },\,\overline{m}_{\mu
})$ are the two real and one complex vector fields, which appear in the
action of the present model. The number of the Poincar\'{e} representations
of the excitation modes may be found looking for the independent variables
of the tetrad $(\ell _{\mu },\,n_{\mu },\,m_{\mu },\,\overline{m}_{\mu })$
which determines the complex structure. It has ($4\times 4=$) 16 real
variables, while four real variables are removed by the extended Weyl
symmetry (\ref{2b7}). Hence we find 12 independent variables. Notice that
this is the number of bosonic modes which exist in the Standard Model, which
has a real massive vector field ($3$ variables), a complex massive vector
field ($2\times 3$ variables), the photon ($2$ variables) and the scalar
Higgs field ($1$ variable). Apparently these arguments are not conclusive
and a more direct calculation is needed.

\subsection{Conformal and Poincar\'{e} vacua}

A direct consequence of the present definition of energy and momentum is
that all the complex structures which are compatible with the Minkowski
metric have zero energy. They determine vacuum configurations. In the
context of the Grassmannian manifold formulation of complex structures we
see that only the Shilov (characteristic) boundary of the classical domain
or its subsurfaces may be vacua. All the complex structures on the closed $%
S^{1}\times S^{3}$\ are vacuum configurations, which are $SU(2,2)$
symmetric, because the surface is invariant. These vacua will be called
conformal vacua, because they break global $SL(4,C)$ symmetry group down to $%
SU(2,2)$ symmetry.

The complex structures on the open "real axis" of the Shilov boundary break
the conformal $SU(2,2)$ symmetry down to the $[Poincar\acute{e}]\times
\lbrack dilatation]$ group. Recall that the \textquotedblleft real
axis\textquotedblright\ subsurface is characterized by a point of the closed
Shilov boundary, which fixes the Cayley transformation (\ref{4d4}). It is
the point of the characteristic boundary in the bounded realization, which
is sent to \textquotedblleft infinity\textquotedblright\ in the unbounded
realization of the classical domain. A general theorem, valid for all
classical domains, states that the automorphic analytic transformations,
which preserve a point of the characteristic boundary in the bounded
realization, become linear transformations in the unbounded realization of
the classical domain\cite{PIAT1966}. In the present case of the $SU(2,2)$
classical domain these linear transformations form the $[Poincar\acute{e}%
]\times \lbrack dilatation]$ group. This argument demonstrates that the
"real axis" vacuum surface breaks global $SL(4,C)$ down to the $[Poincar%
\acute{e}]\times \lbrack dilatation]$ group. One may understand the above
theorem looking at the following general form of the Cayley transformation
which transforms the upper half-plane realization of the $SU(2,2)$ classical
domain onto its bounded realization 
\begin{equation}
z=U_{0}\left( MrM^{\dag }-N^{\dag }\right) \left( MrM^{\dag }-N\right) ^{-1}
\label{4h1}
\end{equation}%
where $\det M\neq 0$\ , $i(N^{\dag }-N)$\ is negative definite and $U_{0}$
is the point of the Shilov boundary, which is sent to infinity. This clear
cut emergence of the Poincar\'{e} group, through a symmetry breaking
mechanism, makes the present model physically very interesting. Recall that
the asymptotic flatness condition generates the BMS group, which does not
appear in Particle Physics.

The $SU(2)$\ $\times U(1)$ transformation $z^{\prime }=Uz$\ changes the
characteristic point $U_{0}$ of the Cayley transformation (\ref{4h1}) to $%
UU_{0}$,\ while it does not affect the Poincar\'{e} transformation. That is,
it changes the Minkowski spacetime, while it does not change the explicit
form of the Poincar\'{e} transformation. This implies that the $SU(2)$\ $%
\times U(1)$ transformation commutes with the Poincar\'{e} transformation in
the following sense: [First make a $U_{0}$ preserving Poincar\'{e}
transformation and after an \textquotedblleft internal\textquotedblright\ $U$
transformation] =[First make an \textquotedblleft
internal\textquotedblright\ $U$ transformation and after a $UU_{0}$
preserving Poincar\'{e} transformation]. I want to point out that these two
subgroups of $SU(2,2)$\ do not commute in the ordinary sense. I think that
this clear cut emergence of the Poincar\'{e} group and the \textquotedblleft
internal\textquotedblright\ $SU(2)$\ $\times U(1)$ group may have some
physical relevance.

The dilatation symmetry is broken by the parameter $a$ of the static Kerr
polynomial which will be described in the next section. But this proof will
be presented here because of the importance of the Poincar\'{e} group in
physics.

In the next section we will see that the first soliton family is generated
by a quadratic polynomial $A_{mn}Z^{m}Z^{n}=0$ with%
\begin{equation}
A_{mn}=%
\begin{pmatrix}
\omega _{AB} & p_{A}^{\quad B^{\prime }} \\ 
p_{\quad B}^{A^{\prime }} & 0%
\end{pmatrix}
\label{4h2}
\end{equation}%
where $p^{AA^{\prime }}=$ $\epsilon ^{AB}p_{B}^{\quad A^{\prime }}.$ This
form is determined assuming invariance under the Poincar\'{e}
transformations $%
\begin{pmatrix}
B_{11} & 0 \\ 
B_{21} & B_{22}%
\end{pmatrix}%
$, with $B_{11}^{\dagger }B_{22}=I$, $\det (B_{22})=1$ and $B_{11}^{\dagger
}B_{21}+B_{21}^{\dagger }B_{11}=0$. If we try to impose dilatation symmetry,
which has the form $%
\begin{pmatrix}
e^{-\rho } & 0 \\ 
0 & e^{\rho }%
\end{pmatrix}%
$, we find $\omega _{AB}=0.$ Apparently this makes the complex structure
trivial. Hence $\omega _{AB}$, which generates the spin of the static
soliton, breaks the dilatation symmetry leaving the Poincar\'{e} group as
the largest symmetry.

\section{"LEPTONIC"\ SOLITONS}

\setcounter{equation}{0}

Standard Model provides a description of weak and electromagnetic
interactions through the classification of the left-handed and right-handed
field components in the representations of the $U(2)$ group. The electron,
muon and heavy lepton (tau) doublets are trivially repeated without any
apparent reason. This is the well known family puzzle. No theoretical
explanation exists of this dummy repetition of only three representations of
the $U(2)$ group. The appearance of the same representations for the quarks
obscures the situation. The extension of the unitary group has not yet
provided any experimentally acceptable model. The present solitonic model
provides a new way to look for a solution to this problem.

\subsection{Three families of solitons}

The dynamical variables of the present model are the gauge field $A_{j\mu
}(x)$ and the integrable tetrad $(\ell _{\mu },\,n_{\mu },\,m_{\mu },\,%
\overline{m}_{\mu })$ up to the extended Weyl symmetry. Notice that the
field equations have the characteristic property to admit pure geometric
solutions with $A_{j\mu }(x)=0$. These are the complex structures defined by
the integrable tetrad or equivalently a rank 2 matrix $X^{mi}(x)$. The
integrable tetrad permits us to define the symmetric tensor (\ref{4p1}),
relative to which the tetrad $(\ell _{\mu },\,n_{\mu },\,m_{\mu },\,%
\overline{m}_{\mu })$ is null. This symmetric tensor may be used as a
metric, which contains a large information from the complex structure. It
has been pointed out the essential difference between the Euclidean complex
structures and the present Lorentzian ones. In the case of Euclidean complex
structures the metric is somehow independent of the complex structure. But
the Lorentzian complex structure is essentially algebraically fixed by the
metric $g_{\mu \nu }$\ because the spinor dyad $o^{A}$ and $\iota ^{A}$,
which determines the integrable tetrad, satisfy the algebraic equation (\ref%
{2a13}). They are principal directions of the Weyl spinor $\Psi _{ABCD}$.

The cornerstone of the soliton theory is the regularity of the solitonic
configurations. In the present case this is translated to the regularity of
the complex manifold. Therefore the Weyl spinor $\Psi _{ABCD}$ must be
regular and $(o^{A},\ \iota ^{A})\ $are roots of a homogeneous fourth degree
polynomial with regular coefficients. Namely, the complex manifold is a
covering space of $\mathrm{R}^{4}$ with a maximum of four sheets and a
minimum of two sheets. That is the solitons of the present model are
algebraically classified into three classes (families) according to the
number of principal null directions of the Weyl tensor as follows:

\begin{itemize}
\item The fourth degree polynomial (\ref{2a13}) is reduced to the square of
a homogeneous second degree polynomial $(\Phi _{AB}\xi ^{A}\xi ^{B})$. These
are the type D spacetimes which admit a regular complex structure. We will
call it "type D family" of the model.

\item The Weyl tensor has three principal null directions, which are
geodetic and shear free. These are type II spacetimes in the Petrov
classification.

\item In the third class the Weyl tensor has four distinct principal null
directions, which must also be geodetic and shear free. These are type I
spacetimes.
\end{itemize}

We should notice the amazing similarities of these three classes with the
three families of leptons and quarks indicating a completely different
approach to the family problem. In conventional Quantum Field Theoretic
models the solution to this problem was searched in the context of large
simple groups for Grand Unified Theories and supergroups for recent
supersymmetric models. In the present model the proposed solution is
topological, based on the Petrov classification of the spacetimes, which is
well known in General Relativity. The present model shows for the first time
that Quantum Field Theory and General Relativity may be intimately related
without Grand Unified Theories, Supersymmetry, Supergravity, Strings and
Superstrings. The study of the stationary axisymmetric solitons of type D
family in the present section will be in the context of this new point of
view.

\subsection{Massive complex structures of the 1$^{st}$ family}

The charged Kerr metric has been extensively studied. After a mass and
charge multipole expansion it was observed\cite{CART1968},\cite{NEWM1973}
that this spacetime had the correct electron gyromagnetic ratio $g=2$.
Notice that this extraordinary result came out in the context of pure
General Relativity without any reference to Quantum Mechanics or any other
particular assumption. This result triggered many attempts to generate
particles in the context of pure General Relativity without apparent
phenomenological success.

The knowledge of the Poincar\'{e} group permit us to look for stationary
(static) axisymmetric solitonic complex structures, which will be
interpreted as particles of the model with precise mass and angular
momentum. In the case of vanishing gauge field, we may use the general
solutions (\ref{2a9}) to find special solutions. In this case the convenient
coordinates are 
\begin{equation}
z^{0}=u+iU\quad ,\quad z^{1}=\zeta \quad ,\quad z^{\widetilde{0}}=v+iV\quad
,\quad z^{\widetilde{1}}=\overline{W}\ \overline{\zeta }  \label{5b2}
\end{equation}%
where $u=t-r$, $v=t+r$ and $t\in R$, $r\in R$, $\zeta =e^{i\varphi }\tan 
\frac{\theta }{2}\in S^{2}$ are assumed to be the four coordinates of the
spacetime surface. Assuming the definitions 
\begin{equation}
z^{0}=i\frac{X^{21}}{X^{01}}\quad ,\quad z^{1}=\frac{X^{11}}{X^{01}}\quad
,\quad z^{\widetilde{0}}=i\frac{X^{32}}{X^{12}}\quad ,\quad z^{\widetilde{1}%
}=-\frac{X^{02}}{X^{12}}  \label{5b3}
\end{equation}%
we look for solutions which are stable along $s^{\mu }=(1,0,0,0)$. That is
we look for massive solutions such that%
\begin{equation}
\delta X^{mi}=i\epsilon ^{0}[\mathrm{P}_{0}]_{n}^{m}X^{ni}  \label{5b4}
\end{equation}%
where $\mathrm{P}_{\mu }=-\frac{1}{2}\gamma _{\mu }(1+\gamma _{5})$.\ It
implies%
\begin{equation}
\begin{array}{l}
\delta X^{0i}=0\qquad ,\qquad \delta X^{1i}=0 \\ 
\\ 
\delta X^{2i}=-i\epsilon ^{0}X^{0i}\qquad ,\qquad \delta X^{3i}=-i\epsilon
^{0}X^{1i}%
\end{array}
\label{5b5}
\end{equation}%
The above definition of the structure coordinates implies%
\begin{equation}
\begin{array}{l}
\delta z^{0}=\epsilon ^{0}\qquad ,\qquad \delta z^{1}=0 \\ 
\\ 
\delta z^{\widetilde{0}}=\epsilon ^{0}\qquad ,\qquad \delta z^{\widetilde{1}%
}=0%
\end{array}
\label{5b6}
\end{equation}%
and consequently%
\begin{equation}
\begin{array}{l}
\delta u=\epsilon ^{0}\qquad ,\qquad \delta U=0 \\ 
\\ 
\delta v=\epsilon ^{0}\qquad ,\qquad \delta V=0 \\ 
\\ 
\delta \zeta =0\qquad ,\qquad \delta W=0%
\end{array}
\label{5b7}
\end{equation}

This procedure gives stable (time independent) solutions. The little group
relative to the vector $s^{\mu }$ is the $SO(3)$\ subgroup of the Lorentz
group. Therefore we may look for solutions, which are \textquotedblleft
eigenstates" of the z-component of the spin. In this case the homogeneous
coordinates satisfy the following transformations%
\begin{equation}
\delta X^{mi}=i\epsilon ^{12}[\mathrm{\Sigma }_{12}]_{n}^{m}X^{ni}
\label{5b8}
\end{equation}%
where $\mathrm{\Sigma }_{\mu \nu }=\frac{1}{2}\sigma _{\mu \nu }=\frac{i}{4}%
(\gamma _{\mu }\gamma _{\nu }-\gamma _{\nu }\gamma _{\mu })$. That is we have%
\begin{equation}
\begin{array}{l}
\delta X^{0i}=-i\frac{\epsilon ^{12}}{2}X^{0i}\qquad ,\qquad \delta X^{1i}=i%
\frac{\epsilon ^{12}}{2}X^{1i} \\ 
\\ 
\delta X^{2i}=-i\frac{\epsilon ^{12}}{2}X^{2i}\qquad ,\qquad \delta X^{3i}=i%
\frac{\epsilon ^{12}}{2}X^{3i}%
\end{array}
\label{5b9}
\end{equation}%
The above definition of the structure coordinates implies%
\begin{equation}
\begin{array}{l}
\delta z^{0}=0\qquad ,\qquad \delta z^{1}=i\epsilon ^{12}z^{1} \\ 
\\ 
\delta z^{\widetilde{0}}=0\qquad ,\qquad \delta z^{\widetilde{1}}=-i\epsilon
^{12}z^{\widetilde{1}}%
\end{array}
\label{5b10}
\end{equation}%
and consequently%
\begin{equation}
\begin{array}{l}
\delta u=0\qquad ,\qquad \delta U=0 \\ 
\\ 
\delta v=0\qquad ,\qquad \delta V=0 \\ 
\\ 
\delta \zeta =i\epsilon ^{12}\zeta \qquad ,\qquad \delta W=0%
\end{array}
\label{5b11}
\end{equation}

A general solution of (\ref{2a9}), which satisfies these symmetries, is
given by the relations%
\begin{equation}
\begin{array}{l}
U=U[z^{1}\overline{z^{1}}]\qquad ,\qquad V=V[z^{\widetilde{1}}\overline{z^{%
\widetilde{1}}}] \\ 
\\ 
W=W[v-u-i(V+U)]%
\end{array}
\label{5b12}
\end{equation}

A static complex structure is expected to be determined by a Kerr function $%
K(X^{m})$\ globally defined on $CP^{3}$. Chow's theorem states that every
complex analytic submanifold of $CP^{n}$ is an algebraic variety (determined
by a polynomial). Hence the present complex structure will be determined by
a quadratic polynomial invariant under (\ref{5b4}) and (\ref{5b8}), which
turns out to be 
\begin{equation}
Z^{1}Z^{2}-Z^{0}Z^{3}+2aZ^{0}Z^{1}=0
\end{equation}

The asymptotic flatness condition (\ref{4e3}) implies 
\begin{equation}
\begin{array}{l}
U=-2a\frac{z^{1}\overline{z^{1}}}{1+z^{1}\overline{z^{1}}}\qquad ,\qquad V=2a%
\frac{z^{\widetilde{1}}\overline{z^{\widetilde{1}}}}{1+z^{\widetilde{1}}%
\overline{z^{\widetilde{1}}}} \\ 
\end{array}%
\end{equation}

A quite general solution is found if $W\overline{W}=1$ ($V+U=0$). In this
case we have the solution%
\begin{equation}
\begin{array}{l}
U=-2a\sin ^{2}\frac{\theta }{2}\qquad ,\qquad V=2a\sin ^{2}\frac{\theta }{2}
\\ 
\\ 
W=\frac{r-ia}{r+ia}e^{-2if(r)}%
\end{array}
\label{5b13}
\end{equation}%
\ A simple investigation shows that this complex structure is (in different
coordinates) the static solution (\ref{2e1}) found in section II using the
Kerr-Schild ansatz. One may easily compute the corresponding tetrad up to
their arbitrary factors $N_{1}$, $N_{2}$ and $N_{3}$.%
\begin{equation}
\begin{array}{l}
\ell =N_{1}[dt-dr-a\sin ^{2}\theta \ d\varphi ] \\ 
\\ 
n=N_{2}[dt+(\frac{r^{2}+a^{2}\cos 2\theta }{r^{2}+a^{2}}-2a\sin ^{2}\theta \ 
\frac{df}{dr})dr-a\sin ^{2}\theta \ d\varphi ] \\ 
\\ 
m=N_{3}[-ia\sin \theta \ (dt-dr)+(r^{2}+a^{2}\cos ^{2}\theta )d\theta
+i(r^{2}+a^{2})\sin \theta d\varphi ]%
\end{array}
\label{5b14}
\end{equation}

The corresponding projective coordinates are%
\begin{equation}
\begin{array}{l}
r_{0^{\prime }0}=i\frac{X^{21}X^{12}-X^{11}X^{22}}{X^{01}X^{12}-X^{11}X^{02}}%
=\frac{z^{0}+(z^{\widetilde{0}}-ib)z^{1}z^{\widetilde{1}}}{1+z^{1}z^{%
\widetilde{1}}} \\ 
\\ 
r_{0^{\prime }1}=i\frac{X^{01}X^{22}-X^{21}X^{02}}{X^{01}X^{12}-X^{11}X^{02}}%
=\frac{(z^{0}-z^{\widetilde{0}}+ib)z^{\widetilde{1}}}{1+z^{1}z^{\widetilde{1}%
}} \\ 
\\ 
r_{1^{\prime }0}=i\frac{X^{31}X^{12}-X^{11}X^{32}}{X^{01}X^{12}-X^{11}X^{02}}%
=\frac{(z^{0}-z^{\widetilde{0}}+ib)z^{1}}{1+z^{1}z^{\widetilde{1}}} \\ 
\\ 
r_{1^{\prime }1}=i\frac{X^{01}X^{32}-X^{31}X^{02}}{X^{01}X^{12}-X^{11}X^{02}}%
=\frac{z^{\widetilde{0}}+(z^{0}+ib)z^{1}z^{\widetilde{1}}}{1+z^{1}z^{%
\widetilde{1}}}%
\end{array}
\label{5b16}
\end{equation}%
If these projective coordinates become a Hermitian matrix $x_{A^{\prime }A}$%
,\ then the complex structure is compatible with the Minkowski metric.
Otherwise, it is a curved spacetime complex structure. The form (\ref{5b13})
has been chosen such that for $f(r)=0$ the complex structure becomes
compatible with the Minkowski metric.

The soliton form factor $f(r)$\ is expected to be fixed by Quantum Theory,
but we have not yet found the precise procedure. We think that any attempt
to give some physical relevance of the present model may come through the
identification of the effective energy-momentum tensor of the excitation
modes with the bosonic part of the Standard Model energy-momentum tensor. In
this case the form factors $f(r)$ of the solitons may be fixed, assuming the
condition that the solitons are particle-like sources of the excitation
modes. Then the massive static soliton (\ref{2e1}) with spin $S_{z}=ma=\frac{%
h}{2}$, should be identified with the electron. Then it will have (in
natural gravitational units $c=G=1$) mass $m=6.8\times 10^{-59}cm$, $%
a=1.9\times 10^{-11}cm$ and charge $q=1.4\times 10^{-34}cm$. The complex
conjugate complex structure would be the positron.

\subsection{Solitonic features of the massive structures}

In ordinary Lorentzian Quantum Field Theory the vacuum is determined as the
stable state with the lowest energy. Solitons are stable states with finite
energies relative to the vacuum. Their configurations are not smoothly
deformable to vacuum configurations. We have already revealed the existence
of two sets of vacua. The conformally invariant vacua, which are complex
structures defined on the closed Shilov boundary $U(2)$ and the Poincar\'{e}
vacua which are complex structures defined on the open \textquotedblleft
real axis\textquotedblright\ of the unbounded neighborhood. In order to
reveal the solitons of the model, we have to use the periodicity criteria.

Recall that the $\phi ^{4}$-model admits two vacua with $\phi =\pm \frac{\mu 
}{\sqrt{\lambda }}$. It is well known that the vacuum configurations are
periodic, while the soliton configurations are not periodic. This
characteristic difference will be used in the present model. The kink
configuration and its excitations satisfy the boundary conditions $\phi
_{kink}(\pm \infty ,t)=\pm \frac{\mu }{\sqrt{\lambda }}$ and the antikink
configuration the opposite ones. The corresponding energy-momentum charges
are related to the gap of the limit values of the field $\phi (x)$ at $\pm
\infty $.

In the present model the excitation modes and the solitons are 4-dimensional
surfaces of $G_{2,2}$ which admit integrable tangent vectors in pairs $(\ell
,\ m)$ and $(n,\ \overline{m})$. Their essential difference will be on the
periodicity of the complex structures they admit. The vacuum surfaces will
admit periodic complex structures, while the solitonic complex structures
are not periodic on the corresponding surfaces. Therefore we have to specify
the precise compactification of the Minkowski spacetime.

Minkowski spacetime is the Poincar\'{e} vacuum of the model and it has
already been identified with a precise open surface of $G_{2,2}$. It is the
\textquotedblleft real axis" in the unbounded realization of the classical
domain. In the bounded realization of the classical domain, it is an open
part of the characteristic (Shilov) boundary. It is precisely limited by the
\textquotedblleft diagonals" $\tau +\rho =\pi \ ,\ (-\pi \leq \tau -\rho
\leq \pi )$, which is $\mathfrak{J}^{\mathfrak{+}}$, and $\tau -\rho =\pi \
,\ (-\pi \leq \tau +\rho \leq \pi )$, which is $\mathfrak{J}^{\mathfrak{-}}$%
. There is an essential difference between Minkowski spacetime and the other
asymptotically flat spacetimes. Every null geodesic which originates at some
point $A^{-}$ of $\mathfrak{I}^{\mathfrak{-}}$ will pass through the same
point $A^{+}$ of $\mathfrak{I}^{\mathfrak{+}}$. This association permits us
to identify $A^{-}$ with $A^{+}$\ compactifying Minkowski spacetime\cite%
{P-R1984}. Notice that all the complex structures, which are compatible with
the Minkowski metric, are well defined on compactified Minkowski spacetime 
\textrm{M}$^{\mathrm{\#}}$,\ because they smoothly cross $\mathfrak{J}=%
\mathfrak{J}^{\mathfrak{+}}=\mathfrak{J}^{\mathfrak{-}}$. The topology of
the whole spacetime \textrm{M}$^{\mathrm{\#}}$\ turns out to be \textrm{M}$^{%
\mathrm{\#}}\sim S^{3}\times S^{1}$.\ 

In order to avoid any misunderstandings, I want to emphasize that there is
an essential difference between the present analysis and the corresponding
Penrose one. In the present model we deal with complex structures while
Penrose deals with Weyl (conformally) equivalent metrics. The present
equivalence relation is larger than the Penrose one. A typical example is
the Schwarzschild spacetime. It is not Weyl (conformally) equivalent with
Minkowski spacetime and it cannot be metrically compactified, because the
first derivatives of the metric do not smoothly cross $\mathfrak{J}^{+}=%
\mathfrak{J}^{-}$.\ But the complex structure of Schwarzschild spacetime is
compatible with Minkowski spacetime because it is trivial. Therefore it can
smoothly cross $\mathfrak{J}$ and the Schwarzschild spacetime is a trivial
vacuum configuration. In the following example of a static solitonic surface
we will consider the Kerr-Schild spacetime but the proof can be extended to
any stationary axisymmetric spacetime with $f(-r)\neq f(r)$. It will be
shown that the integrable tetrad cannot be smoothly extended across $%
\mathfrak{J}$.\ Therefore $\mathfrak{J}^{+}$ and $\mathfrak{J}^{-}$ cannot\
be identified and this complex structure belongs to a soliton sector. The
proof of this failure goes as follows:

In order to make things explicit the Kerr-Newman integrable null tetrad will
be used as an example. Around $\mathfrak{I}^{\mathfrak{+}}$ the coordinates $%
(u,\ w=\frac{1}{r},\ \theta ,\ \varphi )$ are used, where the integrable
tetrad takes the form

\begin{equation}
\begin{array}{l}
\ell =du-a\sin ^{2}\theta \ d\varphi \\ 
\\ 
n=\frac{1-2mw+e^{2}w^{2}+a^{2}w^{2}}{2w^{2}(1+a^{2}w^{2}\cos ^{2}\theta )}%
[w^{2}\ du-\frac{2(1+a^{2}w^{2}\cos ^{2}\theta )}{1-2mw+e^{2}w^{2}+a^{2}w^{2}%
}\ dw-aw^{2}\sin ^{2}\theta \ d\varphi ] \\ 
\\ 
m=\frac{1}{\sqrt{2}w(1+iaw\cos \theta )}[iaw^{2}\sin \theta \
du-(1+a^{2}w^{2}\cos ^{2}\theta )\ d\theta - \\ 
\qquad -i\sin \theta (1+aw^{2})\ d\varphi ]%
\end{array}
\label{5c7}
\end{equation}%
The physical space is for $w>0$\ and the integrable tetrad is regular on $%
\mathfrak{I}^{\mathfrak{+}}$ up to a factor, which does not affect the
congruences, and\ it can be regularly extended to $w<0$. Around $\mathfrak{I}%
^{\mathfrak{-}}$ the coordinates $(v,\ w^{\prime },\ \theta ^{\prime },\
\varphi ^{\prime })$ are used with

\begin{equation}
\begin{array}{l}
dv=du+\frac{2(r^{2}+a^{2})}{r^{2}-2mr+e^{2}+a^{2}}\ dr \\ 
\\ 
dw^{\prime }=-dw\quad ,\quad d\theta ^{\prime }=d\theta \\ 
\\ 
d\varphi ^{\prime }=d\varphi +\frac{2a}{r^{2}-2mr+e^{2}+a^{2}}\ dr%
\end{array}
\label{5c8}
\end{equation}%
and the integrable tetrad takes the form

\begin{equation}
\begin{array}{l}
\ell =\frac{1}{w^{\prime 2}}[w^{\prime 2}\ dv-\frac{2(1+a^{2}w^{\prime
2}\cos ^{2}\theta )}{1+2mw^{\prime }+e^{2}w^{\prime 2}+a^{2}w^{\prime 2}}\
dw^{\prime }-aw^{\prime 2}\sin ^{2}\theta ^{\prime }\ d\varphi ^{\prime }]
\\ 
\\ 
n=\frac{1+2mw^{\prime }+e^{2}w^{\prime 2}+a^{2}w^{\prime 2}}{%
2(1+a^{2}w^{\prime 2}\cos ^{2}\theta ^{\prime })}[dv-a\sin ^{2}\theta
^{\prime }\ d\varphi ^{\prime }] \\ 
\\ 
m=\frac{-1}{\sqrt{2}w^{\prime }(1-iaw^{\prime }\cos \theta ^{\prime })}%
[iaw^{\prime 2}\sin \theta \ dv-(1+a^{2}w^{\prime 2}\cos ^{2}\theta ^{\prime
})\ d\theta ^{\prime }- \\ 
\qquad -i\sin \theta ^{\prime }(1+aw^{\prime 2})\ d\varphi ^{\prime }]%
\end{array}
\label{5c9}
\end{equation}%
The physical space is for $w<0$\ and the integrable tetrad is regular on $%
\mathfrak{I}^{\mathfrak{-}}$ up to a factor, which does not affect the
congruences, and\ it can be regularly extended to $w>0$. If the mass term
vanishes the two regions $\mathfrak{I}^{\mathfrak{+}}$ and $\mathfrak{I}^{%
\mathfrak{-}}$ can be identified\ and the $\ell ^{\mu }$ and $n^{\mu }$
congruences are interchanged, when $\mathfrak{I}^{\mathfrak{+}}$ $(\equiv 
\mathfrak{I}^{\mathfrak{-}})$ is crossed. When $m\neq 0$\ these two regions
cannot be identified and the complex structure cannot be extended across $%
\mathfrak{I}^{\mathfrak{+}}$ and $\mathfrak{I}^{\mathfrak{-}}$.

\subsection{Hopf invariants of complex structure}

We will now consider a classification of the complex structures defined on
the $S^{1}\times S^{3}$ surface (the Shilov boundary) of $G_{2,2}$. They are
determined by two linearly independent functions $\lambda ^{Ai}(\xi )$ in $%
S^{2}$. That is for any complex structure we have two functions%
\begin{equation}
S^{1}\times S^{3}\rightarrow S^{2}  \label{4m2}
\end{equation}%
It is known that the homotopy group $\pi _{1}(S^{2})$ is trivial but $\pi
_{3}(S^{2})=%
\mathbb{Z}
$. The Hopf invariant is determined using the sphere volume 2-form%
\begin{equation}
\omega =\frac{i}{2\pi }\frac{d\lambda \wedge d\overline{\lambda }}{%
(1+\lambda \overline{\lambda })^{2}}  \label{4m3}
\end{equation}%
which is closed. This implies that in $S^{3}$ there is an exact 1-form $%
\omega _{1}$ such that $\omega =d\omega _{1}$. Then the Hopf invariant of $%
\lambda (x)$ is%
\begin{equation}
H(\lambda )=\int \lambda ^{\ast }(\omega )\wedge \omega _{1}  \label{4m4}
\end{equation}%
In the simple case of a linear polynomial $bZ^{0}+Z^{2}=0$, we have%
\begin{equation}
\lambda (x)=\frac{t-z+ib}{x-iy}\quad ,\quad t=0  \label{4m5}
\end{equation}%
The exact 1-form is%
\begin{equation}
\omega _{1}=\frac{ydx-xdy-bdz}{2\pi (x^{2}+y^{2}+z^{2}+b^{2})}  \label{4m6}
\end{equation}%
and%
\begin{equation}
H(\lambda )=-\frac{b}{|b|}  \label{4m7}
\end{equation}%
where we have integrated over the two Minkowski charts, which cover $S^{3}$
by simply permitting $r\in (-\infty \ ,\ +\infty )$. Notice that the present
spinor $\lambda ^{A}(\xi )$ is a solution\ of the linear Kerr polynomial in
the unbounded realization and its Hopf invariant is its helicity. In the
bounded realization the function which defines the mapping $S^{3}\rightarrow
S^{2}$ is different and its Hopf invariant will be different. A simple
transformation shows that in the present case the corresponding mapping has
zero Hopf invariant.

In the case of the solutions of the quadratic Kerr polynomial%
\begin{equation}
\lambda _{\pm }(x)=\frac{-z+ia\pm \sqrt{x^{2}+y^{2}+z^{2}-a^{2}-2iaz}}{x-iy}
\label{4m8}
\end{equation}%
the Hopf invariant can be computed using its relation to the linking
coefficient of two curves in $S^{3}$ determined by the inverse images $%
\lambda ^{-1}(\lambda _{1})=\{x_{1}^{i}(\rho _{1})\}$ and $\lambda
^{-1}(\lambda _{2})=\{x_{2}^{i}(\rho _{2})\}$. Two general curves are
determined using the Lindquist coordinates $(\rho ,\ \theta ,\ \varphi )$ 
\begin{equation}
x^{i}=(\sin \theta \cos \varphi \ ,\ \sin \theta \sin \varphi \ ,\ \cos
\theta )\rho +a(\sin \varphi \ ,\ -\cos \varphi \ ,\ 0)  \label{4m9}
\end{equation}%
for two different values of $\theta ,\varphi $ and the variable $\rho \in
(-\infty ,\ +\infty )$ in order to cover the whole sphere. Then we know that%
\begin{equation}
H(\lambda )=2\frac{1}{4\pi }\int \frac{\varepsilon
_{ijk}(x_{1}^{i}-x_{2}^{i})dx_{1}^{j}dx_{2}^{k}}{|\overrightarrow{x_{1}}-%
\overrightarrow{x_{2}}|^{3}}  \label{4m10}
\end{equation}%
The two curves can be smoothly deformed to the values $\theta _{1}=0$ and $%
\theta _{2}=\frac{\pi }{2}\ ,\ \varphi _{2}=0$. Then the integral becomes%
\begin{equation}
H(\lambda _{\pm })=\frac{a}{2\pi }\int \int \frac{d\rho _{1}d\rho _{2}}{%
(\rho _{1}^{2}+\rho _{2}^{2}+a^{2})^{\frac{3}{2}}}=\pm \frac{a}{|a|}
\label{4m11}
\end{equation}

The curved complex structures which are smooth deformations of conformally
flat spacetimes will have the same Hopf invariants. This is apparently the
case of all the complex structures derived using the Kerr-Schild ansatz. The
curved complex structures have\ Hopf invariant $\frac{a}{|a|}$\ at $%
\mathfrak{J}^{+}$\ and Hopf invariant $-\frac{a}{|a|}$ at $\mathfrak{J}^{-}$.

\subsection{Massless complex structures of the 1$^{st}$ family}

We will now consider configurations $X^{mi}$ which are covariant along a
null vector $s^{\mu }=(1,0,0,1)$. Then $X^{mi}$ satisfy the relations%
\begin{equation}
\delta X^{mi}=i\frac{\epsilon }{2}[P_{0}+P_{3}]_{n}^{m}X^{ni}  \label{5g1}
\end{equation}%
which imply%
\begin{equation}
\begin{array}{l}
\delta X^{0i}=0\qquad ,\qquad \delta X^{1i}=0 \\ 
\\ 
\delta X^{2i}=0\qquad ,\qquad \delta X^{3i}=-i\epsilon X^{1i}%
\end{array}
\label{5g2}
\end{equation}%
In this case the most general quadratic polynomial, which is invariant under
the above transformations is%
\begin{equation}
(bZ^{0}+Z^{2})Z^{1}=0  \label{5g3}
\end{equation}%
\ Notice that this polynomial determines a singular surface in $CP^{3}$,
which may be considered as the limit of the corresponding massive Kerr
polynomial 
\begin{equation}
\begin{array}{l}
A_{mn}Z^{m}Z^{n}=0 \\ 
\\ 
A_{mn}=%
\begin{pmatrix}
0 & 2s & 0 & -E+p \\ 
2s & 0 & E+p & 0 \\ 
0 & E+p & 0 & 0 \\ 
-E+p & 0 & 0 & 0%
\end{pmatrix}%
\end{array}
\label{5gg3}
\end{equation}%
with energy $E$\ and momentum $p$ in the z-direction in the case of
vanishing mass.

The two solutions are $X^{11}=0$ and $X^{02}=-bX^{22}$. In this case we
cannot use the (\ref{5b3}) definitions of the structure coordinates. Instead
we may use the following structure coordinates 
\begin{equation}
z^{0}=i\frac{X^{21}}{X^{01}}\quad ,\quad z^{1}=-i\frac{X^{31}}{X^{01}}\quad
,\quad z^{\widetilde{0}}=i\frac{X^{32}}{X^{12}}\quad ,\quad z^{\widetilde{1}%
}=\frac{X^{02}}{X^{12}}  \label{5g4}
\end{equation}%
Then they transform as follows%
\begin{equation}
\begin{array}{l}
\delta z^{0}=0\qquad ,\qquad \delta z^{1}=0 \\ 
\\ 
\delta z^{\widetilde{0}}=\epsilon \qquad ,\qquad \delta z^{\widetilde{1}}=0%
\end{array}
\label{5g5}
\end{equation}%
and consequently%
\begin{equation}
\begin{array}{l}
\delta u=0\qquad ,\qquad \delta U=0 \\ 
\\ 
\delta v=\epsilon \qquad ,\qquad \delta V=0 \\ 
\\ 
\delta \zeta =0\qquad ,\qquad \delta W=0%
\end{array}
\label{5g6}
\end{equation}%
This procedure gives stable solutions along the null vector $s^{\mu }$. In
the present case the little group is the $E(2)$-like group with the third
generator being the same as the previously studied massive case with the $%
SO(3)$ little group.

The axial symmetry condition (\ref{5b8}) gives the following infinitesimal
transformations%
\begin{equation}
\begin{array}{l}
\delta X^{0i}=-i\frac{\epsilon ^{12}}{2}X^{0i}\qquad ,\qquad \delta X^{1i}=i%
\frac{\epsilon ^{12}}{2}X^{1i} \\ 
\\ 
\delta X^{2i}=-i\frac{\epsilon ^{12}}{2}X^{2i}\qquad ,\qquad \delta X^{3i}=i%
\frac{\epsilon ^{12}}{2}X^{3i}%
\end{array}
\label{5g7}
\end{equation}%
The above new definition of the structure coordinates implies%
\begin{equation}
\begin{array}{l}
\delta z^{0}=0\qquad ,\qquad \delta z^{1}=i\epsilon ^{12}z^{1} \\ 
\\ 
\delta z^{\widetilde{0}}=0\qquad ,\qquad \delta z^{\widetilde{1}}=-i\epsilon
^{12}z^{\widetilde{1}}%
\end{array}
\label{5g8}
\end{equation}

Using the same coordinates $u,\ v,\ \zeta $ a general complex structure
solution is%
\begin{equation}
\begin{array}{l}
U=U[u,\ z^{1}\overline{z^{1}}]\quad ,\quad V=V[z^{\widetilde{1}}\overline{z^{%
\widetilde{1}}}]\quad ,\quad W=W[u+iU] \\ 
\end{array}
\label{5g9}
\end{equation}%
Notice that these relations are not interconnected and they define a general
solution without additional conditions.

In the case of the invariant Kerr quadratic polynomial (\ref{5g3}) we have $%
X^{11}=0$ and $X^{22}=-bX^{02}$. Then the asymptotic flatness conditions (%
\ref{4e3}) imply%
\begin{equation}
\begin{array}{l}
U=0\quad ,\quad V=bz^{\widetilde{1}}\overline{z^{\widetilde{1}}}\quad ,\quad
W=W[u] \\ 
\end{array}
\label{5g10}
\end{equation}

\subsection{The 2$^{nd}$ and 3$^{rd}$ family solitons may be unstable}

It has already been pointed out that the integrability condition $\Psi
_{ABCD}\xi ^{A}\xi ^{B}\xi ^{C}\xi ^{D}=0$ classifies the complex structures
into those with 2, 3, and 4 algebraic sheets. The complex structure with 2
sheets is the type D family\ and it has been extensively studied in the
previous subsections. The application of the same procedure to the type II
and type I families will show that they may not have stable (static)
configurations. That is we will look for an eigenconfiguration which will be
invariant under time translation and z-rotation, and we will find that they
do not exist.

The starting point is the reasonable assumption that for an asymptotically
flat static spacetime there is a coordinate system such that the Weyl tensor
has to approach a Penrose twistor, that is $\Psi _{ABCD}\simeq f\omega
_{ABCD}$. \ This means that the Kerr function can locally become equivalent
to a quartic polynomial $K(Z)=A_{mnpq}Z^{m}Z^{n}Z^{p}Z^{q}=0$. The same
result is found applying Chow's theorem. The existence of an axially
symmetric configuration implies the existence of quartic polynomials such
that%
\begin{equation}
\delta K(Z)=C\varepsilon ^{12}K(Z)  \label{5i1}
\end{equation}%
This relation can be easily solved. The following five solutions are found:

\begin{enumerate}
\item The first solution has $C=2i$ and it contains only the components $%
Z^{0}$\ and $Z^{2}$. The polynomial $K(Z)$ takes the form%
\begin{equation}
\begin{array}{l}
A_{0000}(Z^{0})^{4}+A_{0002}(Z^{0})^{3}(Z^{2})+A_{0022}(Z^{0})^{2}(Z^{2})^{2}+
\\ 
\quad +A_{0222}(Z^{0})(Z^{2})^{3}+A_{2222}(Z^{2})^{4}=0%
\end{array}
\label{5i2}
\end{equation}%
\ which defines a singular surface in $CP^{3}$.

\item Another solution has $C=-2i$ and it is singular too because it
contains only the components $Z^{1}$\ and $Z^{3}$.

\item The third solution has $C=i$ and the polynomial $K(Z)$ takes the form%
\begin{equation}
\begin{array}{l}
A_{0001}(Z^{0})^{3}(Z^{1})+A_{0003}(Z^{0})^{3}(Z^{3})+A_{0012}(Z^{0})^{2}(Z^{1})(Z^{2})+
\\ 
\quad +A_{0023}(Z^{0})^{2}(Z^{2})(Z^{3})+A_{0122}(Z^{0})(Z^{1})(Z^{2})^{2}+
\\ 
\quad
+A_{0223}(Z^{0})(Z^{2})^{2}(Z^{3})+A_{1222}(Z^{1})(Z^{2})^{3}+A_{2223}(Z^{2})^{3}(Z^{3})=0%
\end{array}
\label{5i3}
\end{equation}

\item The fourth solution has $C=-i$ and $K(Z)$ has a form analogous to the
above with $Z^{0}\Leftrightarrow Z^{1}$and $\ Z^{2}\Leftrightarrow Z^{3}$
interchanged.

\item The final solution has $C=0$ and $K(Z)$ takes the form%
\begin{equation}
\begin{array}{l}
A_{0011}(Z^{0})^{2}(Z^{1})^{2}+A_{0013}(Z^{0})^{2}(Z^{1})(Z^{3})+A_{0033}(Z^{0})^{2}(Z^{3})^{2}+
\\ 
\quad
+A_{0112}(Z^{0})(Z^{1})^{2}(Z^{2})+A_{0123}(Z^{0})(Z^{1})(Z^{2})(Z^{3})+ \\ 
\quad +A_{0233}(Z^{0})(Z^{2})(Z^{3})^{2}+A_{1122}(Z^{1})^{2}(Z^{2})^{2}+ \\ 
\quad +A_{1223}(Z^{1})(Z^{2})^{2}(Z^{3})+A_{2233}(Z^{2})^{2}(Z^{3})^{2}=0%
\end{array}
\label{5i4}
\end{equation}
\end{enumerate}

The stability condition relative to time translation 
\begin{equation}
\delta K(Z)=C\varepsilon ^{0}K(Z)  \label{5i5}
\end{equation}%
can now be applied on the above axially symmetric Kerr polynomials. I find
that the only regular (in $CP^{3}$) surface comes from the fifth case. The
invariant polynomial is%
\begin{equation}
A(Z^{1}Z^{2}-Z^{0}Z^{3})^{2}+B(Z^{0}Z^{1})(Z^{1}Z^{2}-Z^{0}Z^{3})+C(Z^{0}Z^{1})^{2}=0
\label{5i6}
\end{equation}%
which may be written as the product of two quadratic polynomials which give
the type D complex structures. Hence we may conclude that the complex
structures (particles) with 3 and 4 sheets cannot be stable.\newpage

\section{"HADRONIC"\ SOLITONS AND\ CONFINEMENT}

\setcounter{equation}{0}

Quark confinement is actually based on the $SU(3)$ gauge group and the non
proven yet hypothesis that the non-Abelian gauge field interactions produce
a confining potential. The perturbative potential of the ordinary Yang-Mills
action is Coulomp-like $\frac{1}{r}$.\ The ordinary Yang-Mills action also
generates the strong P (CP) problem, because it admits instantons which
permit tunnelling between the gauge vacua. The real vacuum of the model is a 
$\theta $-vacuum which generates a parity violation topological term in the
action. The axion particle solution of this problem is expected to be tested
in a LHC experiment. The present model trivially solves these problems
because its modified Yang-Mills action generates a linear static potential
and it does not have instantons.

The amazing similarity between the quark flavor parameters and the leptons
is also a puzzle. The quarks look like leptons with a \textquotedblleft
color\textquotedblright . The theoretical efforts to solve this quark-lepton
correspondence in the context of Grand Unified Theories have affronted
serious problems with the cosmological proton decay bounds. The present
model provides a different way to approach this problem. It seems to imply
that in some approximation for each \textquotedblleft
leptonic\textquotedblright\ (pure geometric) soliton there should be a gauge
field excited soliton, which must be perturbatively confined because of the
linear static potential.

The variables of the present model are the gauge field $A_{j\mu }(x)$ and
the integrable tetrad $(\ell _{\mu },\,n_{\mu },\,m_{\mu },\,\overline{m}%
_{\mu })$. Notice that the field equations have the characteristic property
to admit pure geometric solutions. These solutions may be replaced back into
the new covariant Yang-Mills equations and find the corresponding gauge
field solutions. The simple solution $A_{j\mu }(x)=0$ corresponds to the
pure geometric \textquotedblleft leptonic\textquotedblright\ solitons
without any gauge field interaction.

In complete analogy to the 2-dimensional kinks, we may quantize around the
soliton complex structure\cite{RAG1999}. Then the gauge field configurations
have an asymptotically linear potential instead of the Coulombian ($\frac{1}{%
r}$). This is a clear indication that these excitation modes cannot exist
free and they must be confined into \textquotedblleft
colorless\textquotedblright\ bound states which remind us the hadrons. These
bound states will be hadronic-like solitons with non-vanishing gauge field
strength which in some approximation look like bound states of the simple
\textquotedblleft leptonic\textquotedblright\ excitations through a linear
potential. That is in the present model picture the quarks could be gauge
field excitations of the leptons and they are perturbatively confined. This
very simple picture could explain the complete correspondence between
leptons and quarks. Apparently in the present context, Standard Model should
be considered as an effective theory, like the phonon Lagrangians in solids
and fluids\cite{VOLO2001}. In order to support the above picture of the
soliton sectors, we will first compute the classical potential implied by
the present action.

In spherical coordinates $(t,\ r,\ \theta ,\ \varphi )$ and in the trivial
(vacuum) null tetrad 
\begin{equation}  \label{int2}
\begin{array}{l}
\ell ^\mu =\left( 1\ ,\ 1\ ,\ 0\ ,\ 0\right) \\ 
n^\mu =\frac 12\ \left( 1\ ,\ -1\ ,\ 0\ ,\ 0\right) \\ 
m^\mu =\frac 1{r\sqrt{2}}\ \left( 0\ ,\ 0\ ,\ 1\ ,\ \frac i{\sin \theta
}\right)%
\end{array}%
\end{equation}
the \underline{dynamical variable} of the gauge field is $\left( r\ \sin
\theta \ m^\mu A_{j\mu }\right) $. Assuming the convenient gauge condition 
\begin{equation}  \label{int3}
\overline{m}^\nu \partial _\nu \left( r\ \sin \theta \ m^\mu A_{j\mu
}\right) +m^\nu \partial _\nu \left( r\ \sin \theta \ m^\mu A_{j\mu }\right)
=0
\end{equation}
the field equation takes the form 
\begin{equation}  \label{int4}
\left( \frac{\partial ^2}{\partial t^2}-\frac{\partial ^2}{\partial r^2}%
\right) \left( r\ \sin \theta \ m^\mu A_{j\mu }\right) =\left[ source\right]
\end{equation}
The dynamical variable apparently gives a linear classical
(time-independent) potential. The other two variables $\ell ^\mu A_{j\mu }$
and $n^\mu A_{j\mu }$ of the gauge field decouple and vanish.

Exactly the same approach can be followed in the static soliton sector. The
dynamical variable is now 
\begin{equation}  \label{int5}
A=\left( \left( r+ia\cos \theta \right) \ \sin \theta \ m^\mu A_{j\mu
}\right)
\end{equation}
and it satisfies the gauge condition 
\begin{equation}  \label{int6}
\left( r-ia\cos \theta \right) \overline{m}^\nu \partial _\nu A+\left(
r+ia\cos \theta \right) m^\nu \partial _\nu \overline{A}=0
\end{equation}
The corresponding linear part of the field equation is more complicated but
in the asymptotic limit coincides with (\ref{int4}). The other variables of
the gauge field are r-independent and decouple.

The emergence of the asymptotically linear classical potential implies that
the gauge field modes cannot exist free. They must be confined. The gauge
field excitations of the pure geometric solitons will also be confined
because of the linear potential. An $SU(N)$ gauge group implies that in some
approximation there should be $N$ gauge field excitation modes.\ These
states could look like the three colored quarks. That is, $N$ must equal
three and the gauge group becomes $SU(2)$. Notice that this mechanism
implies the existing in nature correspondence between \textquotedblleft
leptons\textquotedblright\ and \textquotedblleft quarks\textquotedblright .
Namely, for each pure geometric soliton there must be three
\textquotedblleft colored\textquotedblright\ structures which cannot exist
free because of their linear interaction. The confining potential imposes
that that the gauge field excitations cannot exist free. But the existence
of  \textquotedblleft colorless\textquotedblright\ solitons with non
vanishing gauge field gauge field configurations has to be proved. These
solitons are expected to be described by complicated configurations of the
tetrad and gauge field configurations, which satisfy the complicated field
equations of the present action.

We will now show that the Euclidean form of the present Yang-Mills action
does not admit finite action solutions which are called instantons and
measure the tunnelling between the gauge vacua. The proof is based on the
fact that in the Euclidean manifolds, the complex structure becomes the
ordinary real one with $z^{\widetilde{\alpha }}=\overline{z^{\alpha }}$.
Then the structure coordinate form (\ref{2b5}) of the action becomes 
\begin{equation}
\begin{array}{l}
I_{G}=2\int d^{4}\!z\ F_{\!j01}\overline{F_{\!j01}} \\ 
\\ 
F_{j_{01}}=\partial _{0}A_{j1}-\partial _{1}A_{j0}-\gamma
\,f_{jik}A_{i0}A_{k1}%
\end{array}
\label{int7}
\end{equation}%
which is invariant under a complex gauge transformation $A_{j\alpha
}^{\prime }=A_{j\alpha }+\partial _{\alpha }\Lambda _{j}+\gamma
f_{jik}\Lambda _{i}A_{k\alpha }$ , where $\Lambda _{j}$\ are now N complex
functions. Assuming the enlarged gauge condition $A_{j0}=0$ the field
equations become 
\begin{equation}
\begin{array}{l}
\partial _{\overline{0}}F_{\!j01}=0 \\ 
\\ 
\partial _{\overline{1}}F_{j_{01}}-\gamma \,f_{jik}\overline{A_{i0}}%
F_{\!k01}=0%
\end{array}
\label{int8}
\end{equation}

We see that $F_{\!j01}$\ is an holomorphic function of $z^{0}$. On the other
hand the finite action solutions must satisfy the condition 
\begin{equation}
\begin{array}{l}
F_{\!j01}\overline{F_{\!j01}}\underset{\mid z\mid \rightarrow \infty }{%
\Longrightarrow }0 \\ 
\end{array}
\label{int9}
\end{equation}%
That is $F_{\!j01}$\ must be bounded as a function of $z^{0}.$ But we know
that the constant is the only bounded holomorphic function. Hence the finite
action solutions must have $F_{\!j01}=0$.\ Therefore the present model does
not have instantons.

\bigskip

\end{document}